\documentclass[fleqn,usenatbib]{mnras}

\usepackage{amsmath}

\usepackage[T1]{fontenc}
\usepackage{ae,aecompl}
\usepackage{ragged2e}
\usepackage{hyperref}

\usepackage{graphicx}	
\usepackage{amsmath}	
\usepackage{amssymb}	

\newcommand{\angstrom}{\text{\normalfont\AA}}

\title[The Cold Dust Content of Nearby Galaxies]{The Cold Dust Content of the Nearby Galaxies IC\,5325, NGC\,7496, NGC\,7590, and NGC\,7599}

\author[Swapnil S. et al.]{
Swapnil Singh$^{1,}$\thanks{Current affiliation: Space Astronomy Group, U. R. Rao Satellite Centre, Bangalore, 560037, India},
M.\,L.\,N.\,Ashby$^{2}$\thanks{E-mail: mashby@cfa.harvard.edu}, Sarita Vig$^{1}$\thanks{E-mail: sarita@iist.ac.in}, S. K. Ghosh$^{3}$, T. Jarrett$^{4}$, \newauthor \space T.~M.~Crawford$^{5,6}$, Matthew A. Malkan$^{7}$, M.~Archipley$^{8}$ and  J.\,D.\,Vieira$^{8}$
\\
$^{1}$Indian Institute of Space Science \& Technology, Thiruvananthapuram 695547, Kerala, India\\
$^{2}$Center for Astrophysics $|$ Harvard \& Smithsonian,  60 Garden Street, Cambridge, MA 02138, USA \\
$^{3}$Tata Institute of Fundamental Research, Colaba, Mumbai 400005, India \\
$^{4}$Department of Astronomy, University of Cape Town, Private Bag X3, Rondebosch, 7701, South Africa\\
$^{5}$Department of Astronomy and Astrophysics, University of Chicago, 5640 South Ellis Avenue, Chicago, IL, 60637, USA\\
$^{6}$Kavli Institute for Cosmological Physics, University of Chicago, 5640 South Ellis Avenue, Chicago, IL, 60637, USA\\
$^{7}$Department of Physics and Astronomy, UCLA, Los Angeles, CA 90095-1547, USA\\
$^{8}$Department of Astronomy, University of Illinois at Urbana-Champaign, 1002 West Green Street, Urbana, IL, 61801, USA\\
}

\date{Accepted XXX. Received YYY; in original form ZZZ}

\pubyear{2020}

\begin{document}
\label{firstpage}
\pagerange{\pageref{firstpage}--\pageref{lastpage}}
\maketitle

\begin{abstract}
Star-forming galaxies are rich reservoirs of dust, both warm and cold. But the cold dust emission is faint alongside the relatively bright and ubiquitous warm dust emission. Recently, evidence for a very cold dust component has also been revealed via millimeter/submillimeter photometry of some galaxies. This component, despite being the most massive of the three dust components in star-forming galaxies, is by virtue of its very low temperature, faint and hard to detect together with the relatively bright emission from warmer dust. Here we analyze the dust content of a carefully selected sample of four galaxies detected by {\sl IRAS}, {\sl WISE}, and SPT, whose spectral energy distributions (SEDs) were modeled to constrain their potential cold dust content. Low-frequency radio observations using the GMRT were carried out to segregate cold dust emission from non-thermal emission in millimeter/submillimeter wavebands. We also carried out {\sl AstroSat}/UVIT observations for some galaxies to constrain their SED at shorter wavelengths so as to enforce energy balance for the SED modeling. We constructed their SEDs across a vast wavelength range (extending from ultraviolet to radio frequencies) by assembling global photometry from {\sl GALEX} FUV+NUV, UVIT, Johnson BRI, 2MASS, {\sl WISE}, {\sl IRAC}, {\sl IRAS}, {\sl AKARI}, {\sl ISOPHOT}, {\sl Planck} HFI, SPT, and GMRT. The SEDs were modeled with CIGALE to estimate their basic properties, in particular to constrain the masses of their total and very cold dust components.  Although the galaxies’ dust masses are dominated by warmer dust, there are hints of very cold dust in two of the targets, NGC\,7496 and NGC\,7590.
\end{abstract}




\section{INTRODUCTION}\label{introduction}
Dust grains play an important role in the heating and cooling of the interstellar medium (ISM; \citealp{1995ApJ...443..152W}). They are responsible for absorbing nearly half of the total stellar radiation over the lifetime of the Universe \citep{2005ARA&A..43..727L}. This absorbed energy is then re-emitted as thermal radiation at far-infrared (FIR) and submillimeter (submm) wavelengths \citep{1995ApJ...443..152W}. Dust enmeshed in cold, dense molecular clouds plays a significant role in star-formation, and is in some ways a better probe of the cold ISM than the low-lying lines of $^{12}$CO or molecular lines seen at high optical depths \citep{1997A&A...328..471I, 2014ApJ...783...84S}. Detailed observations of dust in the Milky Way (MW) and other galaxies have thus far revealed only an incomplete picture of the physical and chemical properties of dust. Thus, our lack of knowledge of the humble dust grains limits our understanding of their role in star-formation and galaxy evolution \citep{2003ssac.proc...37L}.

\begin{table*}
\begin{center}
\caption{Basic parameters for the sample of galaxies used in this work, taken from NASA/IPAC Extragalactic Database (NED).}
\label{table:galaxies}
\begin{tabular}{c c c c c c c c c}
\hline
\hline
\textbf{Galaxy} & \textbf{RA, Dec (J2000)} & \textbf{z} & \textbf{Distance}& \textbf{Diameter}  &\textbf{Morphology} & \textbf{Activity} &\textbf{SFR} &\textbf{Log M$_*^e$}\\
                & \textbf{(hh:mm:ss, dd:mm:ss)}     & & \textbf{(Mpc)} &\textbf{(arcmin)} &   & \textbf{Type}& \textbf{(M$_{\boldsymbol{\odot}}$yr$^{-1}$)} & \textbf{(M$_{\boldsymbol{\odot}}$)}\\\hline
IC\,5325  & $23$:$28$:$43.43$, $-41$:$20$:$00.5$ & $0.00501$ & $17$& $2.8$& SAB(rs)bc & - & 0.31$^a$, 0.41$^b$  &10.12\\
NGC\,7496 & $23$:$09$:$47.29$, $-43$:$25$:$40.6$ & $0.00550$ & $19$& $3.3$& SB(s)b & Seyfert 2 & 0.81$^c$, 1.35$^c$, & 9.94\\
	     & 		                                  & 	      &        & 	  & 	   & 		   & 2.20$^c$, 1.73$^c$, &\\
	     & 		                                  & 	      &        & 	  & 	   & 		   & 1.97$^c$, 0.31$^b$ &\\
NGC\,7590 & $23$:$18$:$54.81$, $-42$:$14$:$20.6$ & $0.00526$ & $18$& $3.4$& SA(rs)bc & Seyfert 2 &6.20$^d$ & 10.08\\
NGC\,7599 & $23$:$19$:$21.14$, $-42$:$15$:$24.6$ & $0.00551$ & $19$& $4.4$& SA(s)c & - & &10.03\\\hline
\end{tabular}
\end{center}
\justify
$^a$\cite{2007ApJS..173..538T}, $^b$\cite{2007ApJ...658.1006M}, $^c$From different methods described in \cite{2006ApJ...643..173S}, $^d$\cite{1983ApJ...272...54K}, $^e$\cite{2019ApJS..245...25J} 
\end{table*}

FIR/submm emission from dust strongly depends on the properties of the grains and the environment where they reside \citep{2000ApJ...528..799W, 2001ApJ...551..277M}. The FIR output tends to peak at $\sim$100\,$\mu$m. At submm wavelengths, the Rayleigh-Jeans tail of dust emission becomes optically thin even at high column densities (for e.g. $N(\mathrm{H}_2)\sim 2\times10^{24}\,\mathrm{cm}^{-2}$ at 300\,$\mu$m; \citealp{2014ApJ...783...84S, 2015A&A...574L...5P}). Thus, the total dust content in a galaxy can be estimated for a given dust emissivity per unit mass \citep{2007ApJ...663..866D}. In reality, FIR/submm emission from a star-forming galaxy is produced by multiple components over a continuous range of temperatures. Nonetheless, many studies, such as \cite{1986A&A...155..380C} and \cite{2011A&A...532A..56G}, model the FIR emission beyond $40-50$ $\mu$m with three discrete dust components: (i) Very cold dust ($11-14$\,K) associated with quiescent molecular clouds, (ii) Cold dust ($15-25$\,K) associated with atomic hydrogen, and (iii) Warm dust ($30-40$\,K) associated with ionized gas in low density HII regions and warm molecular gas. There is an additional hot dust component having temperatures up to $\sim$\,2000\,K, which is seen in Active Galactic Nuclei (AGN; \citealp{2013ApJ...764..159L,2015ARA&A..53..365N}). This component, coming from the AGN accretion and powerful shocks, is seen in the near-infrared (NIR) and mid-infrared (MIR) regimes. 

Models such as these, although they are simplifications, are highly successful and are widely used to characterize the physical properties of the galaxies\rq\space dust and ISM \citep{2014ApJ...783...84S, 2014PhR...541...45C, 2014A&A...565A.128C}. The terminology used for dust components varies from author to author \citep{2011A&A...532A..56G, 2012ApJ...745...95D, 2013ApJ...768...90L, 2013ApJ...778...51K}. Throughout this paper, we follow the previously mentioned convention of very cold, cold, and warm dust components. 

Local star-forming galaxies serve as a laboratory to study star-formation properties of galaxies that drive the evolution of the galaxy. Star-forming galaxies generate large ultra-violet (UV) fluxes, due to the presence of young O and B type stars, which are absorbed and then re-radiated in the infrared (IR) by the dust. Up to 90\% of a galaxy\rq s UV emission can be shifted into the FIR regime in this way \citep{2010MNRAS.409L...1B}. Thus, the total dust mass in a star-forming galaxy can be used as a tracer of the young stellar population as well as the star-forming potential of the galaxy \citep{1963ApJ...138..393G, 2004ARA&A..42..119V}. Warm dust primarily traces the recently formed young massive stars while cold dust traces the dense molecular clouds from which stars form. Very cold dust emission peaks in the millimeter/submillimeter (mm/submm) regime and the coldest dust component dominates the dust mass despite its small contribution to the total FIR emission \citep{2013MNRAS.433..695C}. 

For example, \cite{1989ApJ...337..650S} showed that the emission at wavelengths longer than 12\,$\mu$m contributes to about one-third of the total flux in some spiral galaxies in the Virgo cluster. Several other, more recent studies have also examined the relation between cold dust and star-formation. \cite{2010A&A...518L..55G} found that the cold and warm dust emission in NGC\,6822 are correlated and hence the cold dust emission also traces star-formation, in addition to the warm dust. \cite{2013MNRAS.433..695C} argue that the temperature of the cold dust component is weakly correlated with the ratio of star-formation rate (SFR) and the dust mass. As a consequence, modeling the dust emission for a galaxy can be used to decipher the SFR in the galaxy.   

Various FIR/submm observatories such as the {\sl Spitzer Space Telescope} \citep{2004AAS...204.3301W}, {\sl AKARI} \citep{2007PASJ...59S.369M}, {\sl Herschel} \citep{2010A&A...518L...1P},  {\sl Planck} \citep{2011AAS...21724301P}, and the South Pole Telescope (SPT; \citealp{2011PASP..123..568C}) have made it possible to investigate the properties of cold dust in nearby galaxies \citep{2000immm.proc...81H, 2000A&A...358...65T, 2004AAS...20514108M, 2008AJ....136.2083M, 2010MNRAS.409....2R, 2013A&A...555A.128T}. There are tantalizing hints in the {\sl Herschel} imaging for the presence of a very cold dust component in nearby galaxies such as NGC\,1140, NGC\,4826, NGC\,7674, NGC\,7793, and the KINGFISH sample \citep{2011A&A...532A..56G, 2014MNRAS.439.2542G, 2012ApJ...745...95D}. But the evidence for the presence of this component so far remains ambiguous \citep{2014MNRAS.441.1040R, 2010A&A...523A..20B}. \cite{2015A&A...574L...5P} used FIR/submm emission across L183, a cold dark cloud, to demonstrate that very cold dust exists but is difficult to identify solely by its emission. This may lead to an underestimation of the dust mass. Some authors also report an anomalous dust emission in the microwave regime ($\sim$\,$10$$-$$90$\,GHz) produced possibly due to rapidly rotating very small dust grains having a non-zero electric dipole moment \citep{2010ApJ...709L.108M, 1998ApJ...508..157D}. Another major challenge is the removal of the potential contamination by thermal and non-thermal radio emission from the very cold dust in the mm/submm regimes \citep{1992ARA&A..30..575C,2010A&A...523A..20B}.

In star-forming galaxies, thermal free-free radiation originates in HII regions and is optically thin. Non-thermal synchrotron radiation, on the other hand, is emitted by relativistic electrons accelerated in the galactic magnetic fields and is usually modeled as a power-law. Although modeling the non-thermal emission is a complex process, simple approximations can be used to estimate the SFRs \citep{1992ARA&A..30..575C}. While the obtained SFRs are consistent with the available values from other SFR indicators, better estimates of supernova rates are needed to strengthen the relationship between the non-thermal emission and SFR. As this non-thermal emission contributes at mm wavelengths, a reliable spectral energy distribution (SED) modeling scheme is essential to obtain precise measurements of masses of the total and very cold dust components in a star-forming galaxy. 

Towards this objective, we selected four low-redshift southern star-forming galaxies based on the criteria discussed in Section\,\ref{sample}. Our observations are presented in Section\,\ref{obs} and the results are discussed in Section\,\ref{result}. The construction of SEDs and their modeling is presented in Section\,\ref{sed}. The nature of evidence of a very cold dust component in these galaxies is discussed in Section\,\ref{vcd}. Finally, the results are summarized in Section\,\ref{summary}.

\section{THE PILOT SAMPLE}\label{sample}
The estimation of accurate dust masses entails submm and mm observations because the warm dust emission dominates at shorter wavelengths \citep{2011A&A...532A..56G,2012MNRAS.419.1833B}. For this work, we therefore selected low-redshift galaxies which were observed by the SPT in three broad bands at frequencies corresponding to 1.4, 2.0, and 3.2\,mm. In order to obtain significant detections, we chose galaxies with flux densities $>$\,10\,mJy in the 1.4\,mm SPT band. We restricted the sample to galaxies having spectral indices $\alpha$ below 1.66, where $\alpha$ is defined through the relation $S_\nu$\,$\propto$\,$\nu^\alpha$ and measured between the SPT 1.4 and 3.2~mm bands to ensure that synchrotron emission does not dominate their submm emission.  We also imposed a requirement that the targets be detected by both {\sl IRAS} and {\sl WISE} so that the warm dust component can be modeled reliably. A size constraint was applied such that the galaxy size along the major axis $<5$\arcmin\ in the {\sl WISE}-22\,$\mu$m maps, for the reliable recovery of their fluxes in the SPT bands. These galaxies were detected by the Sydney University Molonglo Sky Survey (SUMSS; \citealp{2003MNRAS.342.1117M}) at 843\,MHz which provides useful constraints on their non-thermal radio emission. 

A pilot sample of four galaxies visible from the Giant Metrewave Radio Telescope (GMRT), which satisfied all these conditions, was constructed: IC\,5325, NGC\,7496, NGC\,7590, and NGC\,7599. Each galaxy is a spiral with low or moderate SFRs ranging between 0.3 to 6\,M$_\odot$\,{\rm yr}$^{-1}$, based on SFR estimators spanning the electromagnetic spectrum, including both continuum and line emission \citep{1983ApJ...272...54K,2006ApJ...643..173S,2007ApJS..173..538T,2007ApJ...658.1006M}. Some of the indicators include UV, H$\alpha$, IR, and radio luminosities \citep{1998ARA&A..36..189K}. All four galaxies belong to the same southern group of galaxies \citep{1993A&AS..100...47G}. Two of the galaxies in the sample, NGC\,7590 and NGC\,7599, also belong to the Grus-Quartet (a group of four interacting galaxies) along with two other members of the larger group, NGC\,7552 and NGC\,7582 \citep{1996ASPC..106..238K,2009AJ....138..295F}. Basic data for our sample of four galaxies are listed in Table\,\ref{table:galaxies}.

\scriptsize
\begin{table*}
\begin{center}
\caption{Summary of GMRT observations.}
\label{table:obs}
\vspace{2mm}
\begin{tabular}{c c c c c c c c c}
\hline
\hline
\textbf{Galaxy}   & \textbf{Frequency} & \textbf{Date of} & \textbf{On-source }& \textbf{Synthesized} &\textbf{PA} &\textbf{RMS} & \textbf{Peak flux} & \textbf{Integrated}\\ 
   & (MHz) & \textbf{observation} & \textbf{time} (mins)&\textbf{Beam}&(deg)  &(mJy/beam) & (mJy/beam) &\textbf{flux} (mJy)\\ \hline
IC\,5325        & 325 	& 8 Jan 2018 	& 182 & $19.60^{\prime\prime}\times9.01^{\prime\prime}$ 	& $-16.43$ 	&0.43	& 2.88 	&$91.0\pm 9.4$\\
        		& 610 	& 7 Jan 2018 	& 181 & $11.19^{\prime\prime}\times4.61^{\prime\prime}$ 	& 20.00 	&0.04	& 1.36 	&$66.0\pm5.1$\\\hline
NGC\,7496       & 325 	& 31 Dec 2017 	& 140 & $21.44^{\prime\prime}\times10.28^{\prime\prime}$ 	& 5.91		& 0.20 	& 18.99 &$58.7\pm5.0$ \\
       			& 1300 	& 30 Dec 2017 	& 133 & $5.33^{\prime\prime}\times2.20^{\prime\prime}$ 		& 13.05		&0.03	& 13.05 &$21.4 \pm 3.1$ \\\hline
NGC\,7590 		& 325 	& 5 Jan 2018 	& 153 & $22.52^{\prime\prime}\times11.05^{\prime\prime}$ 	&$-3.37$  	&0.80	& 13.77	&$148.2 \pm 14.2$\\ 
  				& 610 	& 10 Jan 2010 	& 183 & $10.86^{\prime\prime}\times4.94^{\prime\prime}$ 	&12.54 		&0.08 	& 12.11	&$112.0\pm6.1$\\ \hline
NGC\,7599  		& 325 	& 5 Jan 2018 	& 153 & $22.52^{\prime\prime}\times11.05^{\prime\prime}$ 	&$-3.37$  	&0.80	& 8.31	&$125.8\pm 11.0$\\ 
  				&610 	& 10 Jan 2010 	& 183 & $10.86^{\prime\prime}\times4.94^{\prime\prime}$ 	&12.54 		&0.08 	& 8.25	& $84.5\pm5.3$\\\hline
\end{tabular}
\end{center}
\justify Errors in the flux densities are estimated using the expression given by \cite{2013ApJ...766..114S}, where the uncertainty in the flux calibration of GMRT is taken to be 5\% \citep{2007MNRAS.374.1085L}.
\end{table*}
\normalsize

\section{OBSERVATIONS AND DATA REDUCTION}  \label{obs}
To obtain precise measurements of the cold dust content in galaxies, reliable SED modeling needs to be performed. Multi-wavelength observations across the electromagnetic spectrum allow us to model each component in a galaxy, while enforcing energy balance. Towards this objective, we performed radio and UV observations. The radio observations were performed using the GMRT, allowing us to isolate synchrotron emission from the long-wavelength thermal tail of the dust emission at submm/mm wavelengths which was obtained from SPT observations (See Section\,\ref{gmrt} and Section\,\ref{spt}). We also obtained UV imaging using the Ultraviolet Imaging Telescope (UVIT) in order to constrain UV emission from the galaxies (See Section\,\ref{uvit}). This ensures reliable energy balance and derivation of the galaxies' basic properties. 

\subsection{GMRT Radio Continuum Observations}\label{gmrt}
The synchrotron flux density can contribute significantly to galaxies' mm luminosity. As synchrotron emission is best studied at low frequencies, GMRT is an ideal choice. GMRT \citep{1991CuSc...60...95S} consists of 30 parabolic dishes of 45m diameter each. Of the 30 antennas, 12 are located in a central array within an area of $\sim$\,1\,km$^2$ and 18 are stretched out along three arms in a Y shaped configuration. The shortest and longest baselines are $\sim$\,100\,m and 25\,km, respectively. The number and configuration of the dishes provide a double advantage: high angular resolution as well as the ability to image the extended diffuse radio emission. All four galaxies in the sample were observed using GMRT in December 2017\,$-$\,January 2018. The galaxies were observed at 325\,MHz, 610\,MHz, and 1300\,MHz with a bandwidth of 32\,MHz. 3C48, 3C138, and 3C147 were used as flux calibrators and the selected phase calibrators were 0010$-$418 and 2314$-$449. In addition, we used archival data for NGC\,7590 and NGC\,7599 at 610\,MHz with 16\,MHz bandwidth, taken from the GMRT Online Archive. The details of the observations are presented in Table\,\ref{table:obs}.

The  data  reduction  was carried out using  the  NRAO\rq s  Astronomical  Image  Processing  System  (AIPS).  The data sets were carefully checked for corrupted data (due to bad baselines, radio frequency interference (RFI), non-working antennas,  etc.) using the tasks {\tt TVFLG}, {\tt UVFLG}, {\tt UVPLT}, and {\tt VPLOT}, which were then removed before proceeding further with the analysis. Calibration was carried out using the tasks {\tt CALIB} and {\tt CLCAL}. To increase the signal-to-noise ratio (SNR), the calibrated data were averaged over several channels. These averaged data were checked again for corrupted data and re-calibrated. These calibrated data were cleaned and deconvolved using the task {\tt IMAGR}, which generates a map of the field using the {\tt CLEAN} algorithm. Multiple iterations of the self-calibration process and {\tt IMAGR} were applied to minimize the amplitude and phase errors. To correct for the loss of sensitivity away from the phase center, {\tt PBCOR} was used for applying a primary beam correction to each image. After this step, final images were generated using {\tt FLATN}. The root-mean square (rms) noise and synthesized beam of the final images are listed in Table\,\ref{table:obs}.
 
\subsection{SPT Millimetre Continuum Observations}\label{spt}
The SPT \citep{2011PASP..123..568C} is a 10\,m telescope located at the National Science Foundation Amundsen-Scott South Pole station in Antarctica.  Initially the SPT was configured with a single mm-wave camera, the SPT-SZ receiver, equipped with 960 detectors capable of observing in three bands centered at roughly 1.4, 2.0, and 3.2\,mm with an angular resolution of 1.0, 1.2, and 1.7 arcmin, respectively.  From 2008 to 2011, the SPT carried out the SPT-SZ survey, covering $\sim$\,2500\,deg$^2$ of southern sky in the three submm bands.  The SPT-SZ survey covers a contiguous region from 20$^{h}$ to 7$^{h}$ in right ascension (R.A.) and $-65^\circ$ to $-40^\circ$ in declination, and was mapped to depths such that the $1 \sigma$ point source sensitivity was approximately 4.0, 1.2, and 2.0\,mJy at 1.4, 2.0, and 3.2\,mm respectively \citep{2020arXiv200303431E}. Fig.\,1 in \cite{2013ApJ...779...86S} shows the field locations and extent of the survey. 

Because cold dust emission peaks in the submm/mm regime (Section\,\ref{introduction}), SPT-SZ photometry formed a key part of the selection criteria (see Section\,\ref{sample}), allowing us to select galaxies with significant mm/submm emission. However, our analysis requires a custom treatment of the SPT-SZ data. All previously published SPT data were for sources that were assumed to be unresolved (point-like) at the arcminute angular resolution of the SPT (e.g., \citep{2010ApJ...719..763V, 2013ApJ...779...61M, 2020arXiv200303431E}, and the standard SPT data pipeline treated them appropriately.  By contrast, the sources considered in this work are resolved by SPT, so a different approach was needed.  Specifically, the time-ordered SPT datastream has to be re-analyzed employing methods that accurately accounted for emission falling outside a single SPT beam.  

The approach adopted here was to use carefully reprocessed WISE 22\,$\mu$m (Band 4) imaging \citep{2019ApJS..245...25J} as a template for the surface brightness distribution in the SPT bands.  We expect the reprocessed WISE Band 4 images, with their sensitivity to warm dust, to be a reasonable proxy for the spatial distribution of the colder dust responsible for the bulk of the emission at mm wavelengths, with a relatively small contamination from starlight continuum and emission lines \citep[e.g.,][]{2014A&A...570A..97I}.  Also, WISE has an angular resolution of 12 arcsec in Band 4, much less than those of the SPT beams.   Our templates are constructed by fitting two-dimensional S{\'e}rsic profiles to the WISE 22\,$\mu$m (Band 4) images \citep{2019ApJS..245...25J}. 

To extract the SPT flux densities from the maps, a $30^\prime \times 30^\prime$ cutout is first extracted from the SPT-SZ maps, centered on the best-fit position of the source in the WISE imaging.  A  three- or five-parameter fit is performed on the three cutouts simultaneously, in which the model is the best-fit WISE Band 4 S{\'e}rsic model, convolved with the known filtering kernel and beam for each SPT band. The parameters of the fit are three amplitudes (one for each SPT band) and, in the five-parameter case, offsets in RA and Dec from the best-fit WISE position to account for any residual astrometry errors in the SPT data.  We find no evidence for any such residual error. The fits are performed in Fourier space, and each Fourier mode is weighted by its expected inverse noise squared. The noise model is the sum of instrument noise (assumed to be white, or uncorrelated between map pixels and bands) and anisotropy in the cosmic microwave background (CMB). The pixel-pixel and band-band correlation in the CMB contribution is properly taken into account.  No explicit contribution from the atmosphere is included in the noise model, because the behavior of atmospheric contamination is significantly similar to that of the CMB in these fits, particularly in the very red angular spectrum.  We derive the final parameter covariance from simulated observations.  For each of the sources in this work, we inject a signal into the true SPT maps near the location of the true source, extract the map cutout at the simulated source location, and estimate the flux density of the simulated source in the same manner as with the true sources. The injected signal near each true source is the WISE Band 4 model, convolved with the SPT filtering kernel and beam, and scaled so that the expected signal-to-noise on the amplitude is roughly 10 in each band. We repeat the procedure 400 times for each source. 

We calculate a three-element residual vector for each source and realization, equal to the extracted best-fit flux density minus the known input flux density in each band, and we use the mean of the outer product of that residual vector over the 400 realizations as our covariance matrix.  As expected, the simulation-based uncertainties differ from those calculated using the noise model (by $\sim 60\%$ at 1.4\,mm, and $\sim 20\%$ at 3.2\,mm).  The off-diagonal components of the final covariance matrix are $<10\%$, so we report the flux densities in the three SPT bands in Table\,\ref{table:fluxes} with uncorrelated uncertainties.

\begin{table*}
\begin{center}
\caption{Details of \textit{AstroSat} - \textbf{UVIT observations}.}
\label{table:uvit}
\begin{tabular}{c c c c c r c c}
\hline
\hline
\textbf{Band} & \textbf{Filter} &  \textbf{$\boldsymbol{\mathrm{\lambda_{mean}}}$}& \textbf{$\boldsymbol{\triangle\lambda}$}  &\textbf{Zero Point} & \textbf{Unit Conversion} & \textbf{Exposure Time} \\
                     & & (\AA) &(\AA) &  \textbf{Magnitude}  & ($\times 10^{-15}$) &(sec) \\\hline
FUV & F154W &  1541 & 380 & 17.765 $\pm$ 0.010 &	3.593 $\pm$ 0.040 & 3221\\
FUV & F172M &  1717 & 125 & 16.341 $\pm$ 0.020 &	10.710 $\pm$ 0.160 & 2593\\
NUV & N245M &  2447 & 280 & 18.500 $\pm$ 0.070 &	0.725 $\pm$ 0.004 & 2723\\
NUV & N279N &  2792 & 90 & 16.500 $\pm$ 0.010 &	3.500 $\pm$ 0.035 & 6420\\\hline
\end{tabular}
\end{center}
\end{table*}

\subsection{\textit{\textbf{AstroSat}} - UVIT Ultraviolet Observations}\label{uvit}
Two galaxies in the sample, NGC\,7590 and NGC\,7599, were imaged using the UVIT on-board the {\sl AstroSat}. {\sl AstroSat} is the first Indian Space Observatory \citep{2004PThPS.155..305A,2014SPIE.9144E..1SS} that observes simultaneously in X-rays ($0.3-100$ keV), UV and visible wavelengths. UVIT is an imaging instrument comprised of two co-aligned telescopes in a Ritchey-Chretien configuration, each having an aperture of 375\,mm. The instrument covers a circular field of view of $\sim$\,$28^\prime$ diameter. One of the telescopes observes in FUV ($1300-1800$\,\AA), the other in NUV ($2000-3000$\,\AA) and visible ($3200-5500$\,\AA). Each channel (FUV, NUV and visible) has several selectable filters with narrower passbands. The spatial resolution achieved by UVIT in FUV/NUV and the visible channel is $<$\,$1.5^{\prime\prime}$ and $<$\,$2.2^{\prime\prime}$ respectively.  More details about the instrument and its in-orbit calibrations and performance can be obtained from \cite{2017JApA...38...28T} and \cite{2017AJ....154..128T}. 

NGC\,7590 and NGC\,7599 were observed by UVIT on 30 November 2017, in two FUV filters (F154W and F172M) and two NUV filters (N245M and N279N). The details of the observations and filters are listed in Table\,\ref{table:uvit}. Simultaneous imaging in a neutral density visible filter was used for aspect reconstruction during the post-observation data processing stage. The Level-1 data products made available by the Indian Space Science Data Center (ISSDC/ISRO) were processed using the UVIT Level-2 Pipeline (UL2P) version V6.3. The pipeline corrects for various instrumental effects (spacecraft drifts, jitter, thermal effects, etc). The UL2P gave the images of the sky in the field of view as the final products, with a pixel size of $0.42^{\prime\prime}\times 0.42^{\prime\prime}$. The intensity unit of the images is counts/sec (CPS). The integrated CPS for all the galaxies for each filter were converted to flux densities ($\rm{F_{filter}}$) and AB magnitudes (${\rm m_{AB}}$) using the following relations as per the prescription of \cite{2017JApA...38...28T}:

\begin{equation}
  \begin{aligned}
    \mathrm{F_{filter}} &= \mathrm{CPS} \times \mathrm{UC}\,\,(\mathrm{erg}\,\,\mathrm{s}^{-1}\,\mathrm{cm}^{-2}\,{\angstrom}^{-1}) \\       
    \mathrm{m_{AB}} &= -2.5\,\,\mathrm{log} (\mathrm{CPS}) + \mathrm{ZP}
  \end{aligned}
\end{equation}
\begin{table}
\begin{center}
\caption{Details of extinction and UVIT flux densities for galaxies in our sample.}
\label{table:uvitflux}
\begin{tabular}{cccr}
\hline
\hline
\textbf{Galaxy} & \textbf{Band} & \textbf{Extinction} & \textbf{Flux Density} \\
                &               & (A$_\lambda$)         & \multicolumn{1}{c}{(mJy)}                 \\\hline
NGC\,7590       & F154W         & 0.123                & 8.49 $\pm$ 0.42       \\
                & F172M         & 0.118                & 9.87 $\pm$ 0.49       \\
                & N245M         & 0.114                & 10.99 $\pm$ 0.55       \\
                & N279N         & 0.092                & 13.44 $\pm$ 0.67      \\\hline
NGC\,7599       & F154W         & 0.126                & 8.92 $\pm$ 0.45       \\
                & F172M         & 0.120                & 10.78 $\pm$ 0.54       \\
                & N245M         & 0.116                & 12.70 $\pm$ 0.64      \\
                & N279N         & 0.094                & 17.53 $\pm$ 0.88      \\\hline
\end{tabular}
\end{center}
\end{table}

\begin{figure*}
\includegraphics[scale=4]{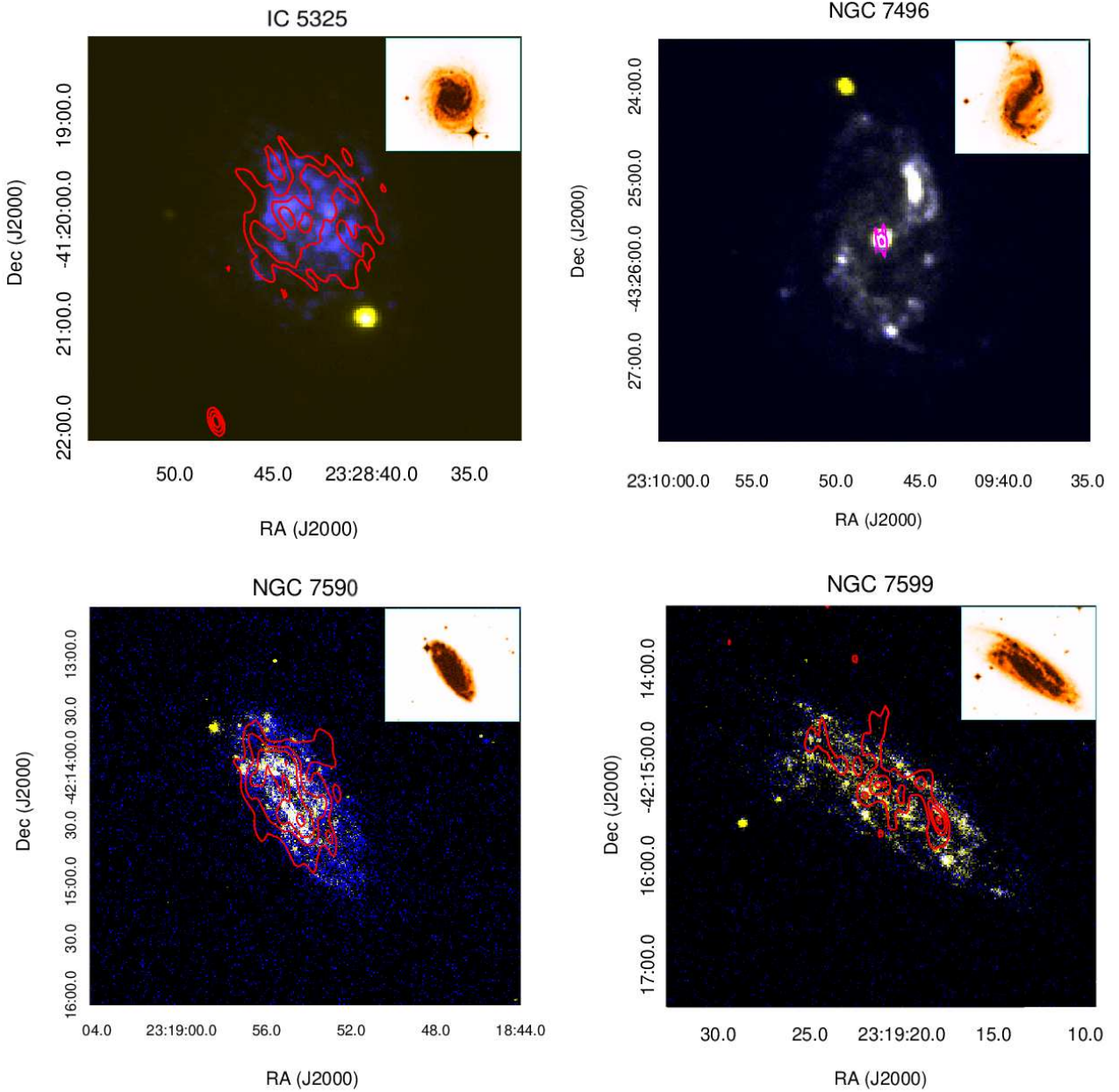}
\caption{Colour composite UV images (NUV: yellow and FUV: blue) for the galaxies in the sample overlaid with GMRT 1300\,MHz (magenta) and 610\,MHz (red) contours. The DSS images for each galaxy ($3.8^\prime\times3.2^\prime$) are shown in the top right corner of each panel. \textit{Top left:} {\sl GALEX} image for IC\,5325 overlaid with 610\,MHz contours. The contour levels are 15, 20, 25 and 30 times $\sigma$ where $\sigma=38\,\mu$Jy/beam. \textit{Top right:} {\sl GALEX} image for NGC\,7496 overlaid with 1300\,MHz contours. The contour levels are 20, 35 and 215 times $\sigma$ where $\sigma=28\,\mu$Jy/beam. \textit{Bottom left:} UVIT image for NGC\,7590 overlaid with 610\,MHz contours. The contour levels are 15, 25, 40 and 45 times $\sigma$ where $\sigma=78\,\mu$Jy/beam. \textit{Bottom right:} UVIT image for NGC\,7599 overlaid with 610\,MHz contours. The contour levels are 15, 25, 40 and 45 times $\sigma$ where $\sigma=78\,\mu$Jy/beam.}
\label{figure:gmrt}
\end{figure*}

\begin{table*}
\begin{center}
\caption{Radio spectral indices for the galaxies in our sample.}
\label{table:spi}
\vspace{2mm}
\begin{tabular}{ccccc}
\hline
\hline
\textbf{Galaxy}      				& \textbf{IC\,5325}   	& \textbf{NGC\,7496}    	& \textbf{NGC\,7590}    		& \textbf{NGC\,7599}         \\\hline
Spectral Index from GMRT 			& $-0.51\pm0.72$ $^a$	& $-0.42\pm 0.14$ $^b$      & $-0.46\pm 0.25$ $^a$          & $-0.63\pm 0.24$ $^a$       \\
Spectral Index from all radio data 	& $-1.04\pm0.32$		& $-0.61\pm 0.12$        	& $-0.64\pm 0.06$               & $-0.62\pm 0.24$            \\\hline
\end{tabular}
\end{center}
\begin{flushleft}
$^a 325-610$\,MHz and $^b 325-1300$\,MHz
\end{flushleft}
\end{table*}

\justify The values of zeropoint magnitudes (ZP) and Unit conversion factors (UC) are listed in Table\,\ref{table:uvit}. The uncertainties in flux densities for each filter are taken as 5\% considering (i)~the uncertainties in CPS arising from Poisson noise of the photon-counting detector, and (ii)~the error in the UC factor from calibration uncertainties \citep{2016ApJ...833L..27S}.

The obtained flux densities have been corrected for foreground extinction in the MW. Adopting the ratio of total-to-selective extinction as $\mathrm{R_{V}}= 3.1$ \citep{1958AJ.....63..201W} for the MW, the extinction coefficient in V band ($\mathrm{A_{V}}$) was calculated for each of the galaxies. Using the reddening relation of \cite{1989ApJ...345..245C}, this $\mathrm{A_{V}}$ was used to estimate and correct for the extinction in each filter, as listed in Table\,\ref{table:uvitflux}.

\section{RESULTS}\label{result}

\begin{figure*}
\begin{center}
\includegraphics[scale=0.4]{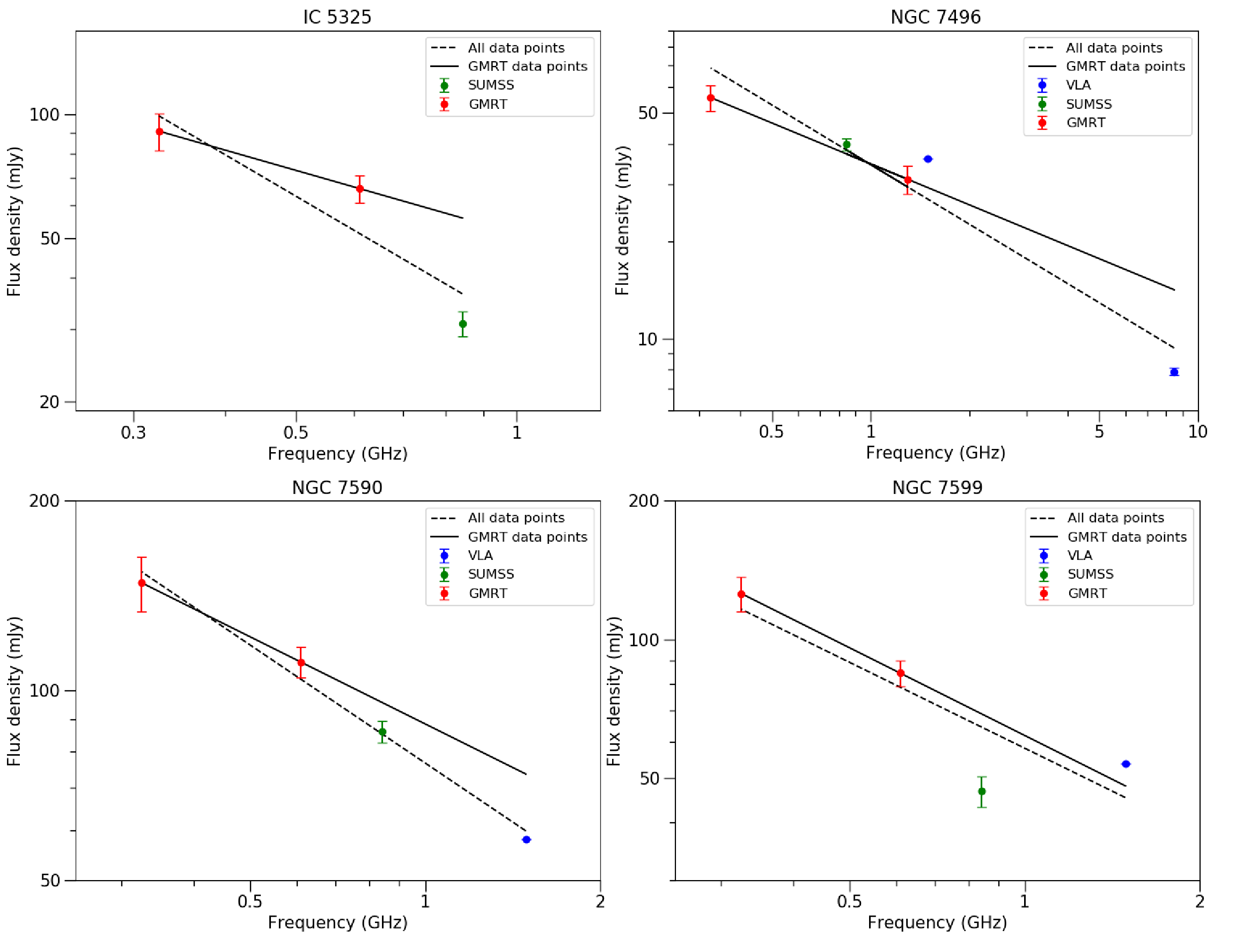}
\end{center}
\caption{Radio-frequency SEDs for the pilot sample. The red, green and blue data points are flux densities from GMRT, SUMSS and VLA, respectively. The solid line indicates the spectral slope calculated from the GMRT data only while the dashed line indicates the spectral slope inferred from all available data.}
\label{figure:spec}
\end{figure*}

\subsection{Radio Emission from the Galaxies}\label{result_radio}
The distribution of radio emission from ionized gas in galaxies can help us understand the emission mechanism and provide hints about the star-formation activity in addition to possible activity from a central black hole. The GMRT has a minimum baseline of $\sim100$\,m and is thus able to detect large structures up to a size of 32$^\prime$, 17$^\prime$ and 9$^\prime$ at frequencies of 325, 610 and 1300\,MHz, respectively. Each galaxy in our sample has a diameter $<5$\,arcmin.  Thus, the GMRT is capable of detecting most of the galaxies\rq\space emission at these frequencies and it is unlikely that any diffuse emission has been missed. The radio emission from our sample of galaxies is shown as contours in Fig.\,\ref{figure:gmrt}. The details of the emission from each galaxy are discussed below:

\justify \textit{\textbf{IC\,5325:}} The 325 and 610\,MHz maps suggest that the emission is spread over the disk with the peak emission being close to the center of the galaxy. 
\\\textit{\textbf{NGC\,7496:}} The emission at 325\,MHz is distributed across the galaxy with the peak emission at the center of the galaxy. We note that the emission traces one of the spiral arms in the galaxy. The 1300\,MHz emission is detected solely towards the nucleus of the galaxy, which might be due to the presence of an AGN. 
\\\textit{\textbf{NGC\,7590:}} The emission at both frequencies (325 and 610\,MHz) is distributed throughout the galaxy.
\\\textit{\textbf{NGC\,7599:}} We observe that the emission at both 325 and 610\,MHz peaks in one of the spiral arms of the galaxy. This region outshines the rest of the galaxy as not much emission is seen from the other regions in the galaxy. This may indicate some intense star-formation activity happening in that spiral arm of the galaxy. Alternately, it is possible that this is an effect of superposition of emission from an alternate radio source in the sky. But this hotspot also lies towards the HI \lq bridge\rq \space linking NGC\,7599 and NGC\,7590 as seen by \cite{2009AJ....138..295F}. This implies that the two galaxies are interacting and argues against the chance-superposition hypothesis. We also see a high UV flux in the region, which is indicative of young star-formation (see Section\,\ref{UVem}).

We combined the GMRT detections with all other available radio flux measurements to construct the radio SEDs of these galaxies. The VLA flux densities were obtained from NED, while the 843\,MHz flux densities from SUMSS \citep{2001ASPC..232...53S} were extracted from the images obtained from their archive\footnote{http://www.astrop.physics.usyd.edu.au/cgi-bin/postage.pl}. We estimated the spectral indices $\alpha$ adopting the standard definition (See Section \ref{sample}). The value of $\alpha=-0.1$ indicates thermal optically thin emission, while for synchrotron optically thin emission typically $\alpha=-0.8$. The spectral indices for the galaxies were calculated using two methods: (i) Using GMRT flux densities alone, and (ii) using all available flux measurements from multiple interferometric observations. The former was carried out by restricting the GMRT image visibilities to the same {\sl uv} range ($\sim$\,$95\lambda$ to $23$k$\lambda$) so that fluxes within the same spatial scales are considered. In the latter method, the flux densities from all available images were used although they were taken with different array configurations. As a consequence, the spectral indices from the latter method can be considered approximate at best.  The spectral indices are listed in Table\,\ref{table:spi} and shown as lines in Fig.\,\ref{figure:spec}.

\begin{table*}
\begin{center}
\caption{Estimated SFRs (in $\mathrm{M}_\odot$\,{\rm yr}$^{-1}$) and sSFRs (in Log {\rm yr}$^{-1}$) for the sample galaxies using various indicators. (Column\,2) SFR is based on a monochromatic {\sl IRAS} 60\,$\mu$m flux converted to SFR using the conversion relation prescribed by \protect\cite{2008ASPC..381..216R}. (Column\,3) SFR computed using FIR luminosity following the method given by \protect\cite{2002AJ....124.3135K}. (Column\,4) SFR estimated from the GMRT 610\,MHz flux using the relation given by \protect\cite{2009MNRAS.397.1101G}. (Column\,5) SFR computed from the UVIT F154W and {\sl GALEX} FUV absolute magnitudes using the method given by \protect\cite{2013AJ....146...46K}. (Column\,6) SFR computed from the SED modeling as described in Section\,\ref{bestfitseds}. (Column\,7) SFR estimated using the hybrid estimator $-$ FUV and 25\,$\mu$m fluxes, following the relations given by \protect \cite{2011ApJ...741..124H}. (Columns\,8-11) SFRs and specific SFRs using the {\sl WISE} W3 and W4 bands were obtained from the {\sl WISE} Extended Source Catalogue by \protect\cite{2019ApJS..245...25J}.}   
\label{table:sfrcalc}
\scriptsize
\begin{tabular}{l c c c c c c c c c c}
\hline
\hline
\textbf{Galaxy}  &  \textbf{60\,$\boldsymbol{\mu}$m $^a$} &	\textbf{FIR$^a$} & \textbf{610\,MHz $^a$} & \textbf{FUV $^a$} & \textbf{SED} &  \textbf{FUV$\mathbf{+25}$\,$\boldsymbol{\mu}$m } &\textbf{W3} &\textbf{W4}  & \textbf{sSFR$_{\boldsymbol{\mathrm{W3}}}$} & \textbf{sSFR$_{\boldsymbol{\mathrm{W4}}}$}\\\hline
IC\,5325  		& $0.69 \pm 0.05$ 	&$0.93\pm0.05$	& $0.67 \pm 0.05$ 	& $0.44 \pm 0.01$ 	&$2.06\pm0.10$ & $ 1.00\pm 0.05$ & $1.55\pm0.60$		& $1.14\pm0.46$ 	& $-9.93$	& $-10.06$  		\\
NGC\,7496$^*$ 	& $1.95 \pm 0.18$ 	&$1.90\pm0.11$	& - 				& $0.25 \pm 0.01$ 	&$1.19\pm0.06$ & $ 3.13 \pm 0.24$ & $2.25\pm0.88$ 	& $3.76\pm1.51$		& $-9.58$	& $-9.36$			\\
NGC\,7590$^*$ 	& $1.32 \pm 0.13$ 	&$1.61\pm0.10$	& $1.27 \pm 0.07$ 	& $0.47 \pm 0.02$ 	&$2.16\pm0.11$ & $ 1.66 \pm 0.11$ & $2.15\pm0.74$ 	& $1.68\pm0.66$		& $-9.75$	& $-9.86$			\\
NGC\,7599 		& $1.23 \pm 0.14$ 	&$1.54\pm0.10$	& $1.07 \pm 0.07$ 	& $0.55 \pm 0.03$	&$3.39\pm0.40$ & $ 1.70 \pm 0.13$ & $2.28\pm0.88$		& $1.65\pm0.67$		& $-9.68$ 	& $-9.82$			\\\hline
\end{tabular}
\end{center}
\justify \scriptsize$^a$  SFR uncertainties estimated using flux uncertainties alone. $^*$Seyfert type 2 galaxies.
\end{table*}
\normalsize

\subsection{UV Emission from Galaxies}\label{UVem}
The UV images of NGC\,7590 and NGC\,7599 from UVIT are shown in Fig.\,\ref{figure:gmrt}. For the other two galaxies, we have considered the images taken by the Galaxy Evolution Explorer ({\sl GALEX}). In each of the image panels, we have also included the DSS optical images in the top right. The UV images trace the location of young stars in the galaxies, with dominant contribution from OB stars. The optical images, on the other hand, trace the older stellar population. \cite{2012ARA&A..50..531K} suggest that the NUV and FUV emission trace recent star-formation with stars having ages up to 200 and 100\,{\rm Myr}, respectively.  We briefly describe the morphology of these galaxies in the UV regime and compare them with optical. 

In IC\,5325, the {\sl GALEX} UV emission appears to trace the fragmented arms of the galaxy. The emission is mostly from the inner regions of the disk. The galaxy is brighter in FUV as compared to NUV, which implies that recent star-formation has occurred in the galaxy in the last 100\,{\rm Myr}. The optical image for IC\,5325 shows that the inner disk of the galaxy is dominated by late-type stars. In NGC\,7496, we see strong emission in both NUV and FUV from the center of the galaxy as well as clumpy emission from the two spiral arms. The northern spiral arm shows traces of young massive stellar complexes unlike the southern arm, where  regions towards the ends of the arms indicate the presence of young stellar complexes. This suggests that these regions are composed of young stars having ages between 0$-$200\,{\rm Myr}. In this galaxy, the emission from the older stellar population comes from the center of the galaxy as well as the two arms. In NGC\,7590, UV emission traces the spiral arms of the galaxy with more younger stars located in the inner regions of arms close to the center of the galaxy. In this galaxy as well, an equal emission from NUV and FUV is observed indicating stellar ages in the range 0$-$200\,{\rm Myr}. From the optical image of NGC\,7590, we see that the old stellar population is well distributed throughout the disk of the galaxy. For NGC\,7599, the images generated from UVIT observations show that the UV emission traces the spiral arms of the galaxy. A similar trend is seen in the optical image as well. The composite UV image shows that the galaxy is dominated mostly by NUV emission with young stars having an age of up to 200\,{\rm Myr}. From the FUV image of this galaxy, we observe that the same region having high flux in the radio image has marginally higher emission than that from the surrounding regions in the galaxy. The optical image also shows that this region does not have a significant older population as compared to the young stellar population.

\subsection{Integrated Star-formation Rates of the Galaxies}
The SFR of a galaxy constrains its evolutionary history. Most of the methods for SFR estimation are derived from correlations in large galaxy samples. These correlations are not very tight and could introduce an inherent selection bias in the SFR relations. The derived SFRs are usually highly uncertain in galaxies due to lack of information regarding (i) details of various physical processes contributing to star-formation in the galaxy, and (ii) precise estimates of physical conditions in the galaxy. Nevertheless, the SFR of a galaxy provides useful insights regarding its evolution and can be used for comparison of SFRs with other galaxies, derived using similar methods. 

The SFRs of the sample of galaxies from literature are listed in Table\,\ref{table:galaxies} and their values differ depending on the wavelength used. In the current work, we compare the SFRs obtained by employing infrared (60\,$\mu$m), FIR and radio (610\,MHz) flux densities as well as FUV (UVIT/{\sl GALEX}) absolute magnitudes. 

The SFR estimated using the 60\,$\mu$m flux density ($\psi$) is given by the following relation from \cite{2008ASPC..381..216R}.
\begin{equation}
\left( \frac{\psi}{\mathrm{M}_\odot {\rm yr}^{-1}} \right)= 2.2 \epsilon^{-1}10^{-10} \left( \frac{L_{60}}{L_\odot} \right)
\end{equation}
\justify Here, $\epsilon=2/3$ represents the fraction of UV light absorbed by dust. 

We also estimated the galaxies' SFRs using the FIR luminosity following the method prescribed by \cite{2002AJ....124.3135K}:
\begin{equation}
\mathrm{SFR_{IR}\,\,(M_\odot\,yr^{-1})} \sim 7.9 \times 10^{-44}\, \mathrm{L_{FIR}\,\,(ergs\,\,s^{-1})}
\end{equation}
where $\mathrm{L_{FIR}}$ is the far-infrared luminosity computed from the {\sl IRAS} 60\,$\mu$m and 100\,$\mu$m fluxes, and the distance to the galaxy.

At 610\,MHz, \cite{2009MNRAS.397.1101G} provide two ways to estimate the galaxy SFR ($\phi$) depending on galaxy luminosity. This treatment includes the effect of non-thermal emission assuming a spectral index of $-0.8$.
\begin{equation}
\left(  \frac{\phi}{\mathrm{M}_\odot {\rm yr}^{-1}}\right)=2.84 \times 10^{-22} \left(  \frac{L_{610}}{\mathrm{W\,Hz}^{-1}}\right);\,\,  L_{610} > L_c
\end{equation}
\begin{equation}
\left(  \frac{\phi}{\mathrm{M}_\odot {\rm yr}^{-1}}\right)=\frac{2.84 \times 10^{-22}}{0.1+0.9(L_{610}/L_c)^{0.3}} \left(  \frac{L_{610}}{\mathrm{W\,Hz}^{-1}}\right);\,\, L_{610} \leq L_c
\end{equation}
Here $L_c =3.3 \times 10^{21} $W Hz$^{-1}$ is the luminosity at 610 MHz of an $\sim$\,$L_*$ galaxy with $\phi \sim 1$\,M$_\odot {\rm yr}^{-1}$. 
\justify We also estimated the SFRs using the UVIT F154W and {\sl GALEX} FUV absolute magnitudes. The following method prescribed by \cite{2013AJ....146...46K} was used to compute the SFRs:
\begin{equation}
\mathrm{log(SFR[M}_\odot\,\mathrm{yr}^{-1}]) = 2.78 - 0.4m^c_{FUV} + 2 \mathrm{log(D)}
\end{equation} 
Here, $m^c_{FUV}$ is the FUV magnitude corrected for extinction and D is the kinematic distance to the galaxy.

The SFRs computed using various methods are presented in Table\,\ref{table:sfrcalc}. We find that SFR values obtained using radio emission are in the range $0.8$\,$-$\,$1.3$\,M$_\odot$\,{\rm yr}$^{-1}$. The SFR estimates obtained from both FIR and radio indicators are strikingly similar despite the fact that each depends on a number of simplifying assumptions. While deriving the relation for estimating SFRs using the 60\,$\mu$m luminosity, the conversion between $L_{IR}$ and $L_{60}$ has uncertainties in it and the parameter, $\epsilon$, used to calculate the amount of young star-formation could also vary from galaxy to galaxy. The SFR-610\,MHz luminosity relation, on the other hand, bears uncertainties in the conversion of the 1.4\,GHz luminosity to 610\,MHz luminosity due to the assumption of a constant radio spectral index. The direct relation between non-thermal synchrotron emission and star-formation is poorly understood. This is due to our lack of understanding regarding contributions of the intermediate processes, such as relations between supernova remnants and acceleration of electrons to relativistic energies, their propagation and energy loss mechanisms \citep{1992ARA&A..30..575C}. 

Consequently, alternate measures of star-formation such as in the infrared are often used to derive SFR correlations in radio \citep{2002AJ....124.3135K}. As thermal emission is not considered for computing the SFRs in the method by \cite{2009MNRAS.397.1101G}, physical effects such as recent starburst activity or suppression of the radio luminosity in galaxies could also lead to inaccurate estimations of SFRs at radio wavelengths. The SFRs calculated using FUV emission yield the lowest values of SFRs ranging between $0.3$\,$-$\,$0.6$\,M$_\odot$\,{\rm yr}$^{-1}$. We believe this to due to the high susceptibility of FUV to extinction. SFR estimates using FUV emission can be inaccurate due to the internal light extinction in the galaxies, which would be larger in dusty galaxies such as those in our sample and is difficult to estimate. To account for this, we also estimate the SFRs for the galaxies using a hybrid SFR indicator which is a combination of the FUV and the 25\,$\mu$m (MIR) fluxes \citep{2011ApJ...741..124H}. These SFRs range from $1.00-3.13$\,M$_\odot$\,{\rm yr}$^{-1}$. However, we note that the SFR estimates for two of the galaxies which have an AGN are higher than the other indicators, especially for NGC\,7496. This could be due to the contribution from the AGN which is also seen as excess in MIR regime. It challenging to isolate SF activity from the AGN activity and therefore difficult to obtain reliable SFR estimates for these galaxies. We have also listed the SFRs and specific star-formation ratios (sSFR = SFR/mass) from the {\sl WISE} Extended Source Catalogue (WXSC) by \cite{2019ApJS..245...25J} in Table\,\ref{table:sfrcalc} for comparison. \cite{2019ApJS..245...25J} followed the prescription given by \cite{2017ApJ...850...68C} to estimate the SFRs for the galaxies using {\sl WISE} W3 ($12$\,$\mu$m) and W4 ($22$\,$\mu$m) bands. The SFRs derived from the {\sl WISE} W3 and W4 bands are significantly higher than the SFRs estimated from other methods. This method avoids uncertainties due to extinction but can produce inaccurate estimates in the presence of strong silicate absorption features in dusty starburst galaxies as well as powerful AGN. The sSFRs from {\sl WISE} W3 and W4 bands are not significantly different and range from $-9.4$ to $-10.1$ for our sample of galaxies. In general, the SFR estimates for NGC\,7496 and NGC\,7590, derived from various methods, should be interpreted with caution due to the presence of an AGN.

For local galaxies, the SFR ranges from $0.1$\,$-$\,$100$\,$M_\odot$\,{\rm yr}$^{-1}$ depending on the type of the galaxy (normal or starburst), including the MW having an SFR $\sim$\,$1.9 \pm 0.5$\,$\mathrm{M}_\odot$\,{\rm yr}$^{-1}$ \citep{1998ApJ...498..541K, 2011AJ....142..197C}. SFRs up to $\sim$\,$20$\,M$_\odot${\rm yr}$^{-1}$ have been estimated for gas-rich spiral galaxies \citep{1998ARA&A..36..189K}. In our sample, all the galaxies have relatively low SFRs, similar to the MW. The SFRs for the galaxies suggest that all of them are normal star-forming spirals.  \cite{2019ApJS..245...25J} presented a galaxy star-formation main sequence (GMS) diagram, which represents the past-to-present Star Formation History (SFH) for galaxies (See Fig.\,16 of their paper). Based on the sSFR values, NGC\,7590 and NGC\,7599 can be placed at the upper end of the sequence as they are young, dusty galaxies with high stellar masses. By contrast, IC\,5325 seems to have consumed most of its gas to attain its high stellar mass and it falls in the category of intermediate disk galaxies (like the MW and Andromeda) on the GMS diagram. NGC\,7496 also falls in the same category as NGC\,7590 and NGC\,7599, but due to the presence of a strong AGN, one cannot be certain about its MIR derived SFR as well as the stellar mass \citep{2019ApJS..245...25J}, and hence its evolutionary stage.  

\subsection{Triggered Star-formation in NGC\,7599?}\label{trigsf}
NGC\,7599 is a normal star-forming galaxy but shows some unusual features. NGC\,7599 has faint radio emission at the nucleus, but has higher flux densities towards one of the spiral arms \citep{2014MNRAS.437.3236F}. This could be associated with a recent star-formation event and hence we refer to this region as an anomalous star-forming region. This anomalous star-forming region accounts for $\sim$\,$10$\% of the total flux from the galaxy at 610\,MHz. The reason behind such high flux, which is much greater than the rest of the galaxy, is not entirely clear, but some clues are evident in the multi-wavelength imaging. The region is also seen to be bright in the {\sl WISE} imaging indicating the presence of a dusty region, as seen in star-forming regions \citep{2019ApJS..245...25J}. We suspect that this could be due to the interaction with its neighboring galaxy NGC\,7590. This arm lies towards the HI \lq bridge\rq \space found to link NGC\,7599 with NGC\,7590 \citep{2009AJ....138..295F}. Interactions between galaxies can perturb gas within the galaxy and lead to the formation of elongated structures like tails and bridges and can also trigger star-formation in any of the interacting galaxies. The UVIT-FUV image appears to show that the anomalous star-forming region has a flux density which is marginally higher than the emission from the surrounding region in the galaxy. Because FUV imaging indicates the presence of young stars, it is possible that there is some interaction between the two galaxies which is actively triggering star-formation -- but in a localized, intense fashion -- in the spiral arm of NGC\,7599. Intriguingly, this feature does not stand out in the optical or IR images of the NGC\,7599. Further investigation at higher resolution and sensitivity is required in order to better understand the origin of the anamolous star-forming region in this otherwise normal galaxy.

\section{SED MODELING}\label{sed}
The information about various physical processes that take place in the galaxy is embedded in the SED of the galaxy. Various components (stellar, dust, nebular, etc.) contributing to the galaxy luminosity can be extracted. Therefore, a detailed analysis of the SED of a galaxy, with the aid of models, can be used to derive the various physical parameters of the galaxy \citep{1999A&A...350..381D, 2011A&A...527A.109P, 2014MNRAS.440.1880W}. In particular, SED modeling is a useful tool to understand the past and current star-formation history of the galaxy. In addition, SEDs furnish information about the dust luminosities, masses, emissivities etc. and enable us to perform a detailed study of dust emission from the galaxy \citep{2012ApJ...752...55K,2019arXiv190602712L}. The AGN activity in the galaxy can also be deduced from the SEDs. The SED modeling, thus, provides comprehensive information about the physical processes driving the evolution of the galaxy.

\subsection{Multiwavelength Data}
For the SED modeling described in this section, we made use of all available photometry from UV to radio wavelengths. The multitude of photometry (Table\,\ref{table:fluxes}) is necessary to separate the multiple blended and overlapping contributors to the SEDs, and ultimately characterize the galaxies' cold dust content. In addition to the UVIT, GMRT, and SPT continuum measurements described in Section\,\ref{obs}, a considerable quantity of additional photometry was assembled for our modeling effort. This includes global photometry from the {\sl GALEX} FUV+NUV bands \citep{2005ApJ...619L...1M}, Johnson BRI, 2MASS $JHK_s$ \citep{2006AJ....131.1163S}, {\sl WISE} bands from 3 to 22\,$\mu$m \citep{2010AJ....140.1868W}, {\sl IRAC} bands from 3.6 to 8.0\,$\mu$m \citep{2004ApJS..154...10F}, {\sl IRAS} bands from 12 to 100\,$\mu$m \citep{1984ApJ...278L...1N}, {\sl Planck} HFI at 217, 353, 545 and 857\,GHz \citep{2011AAS...21724301P}, and SUMSS at 843\,MHz \citep{2003MNRAS.342.1117M}. The flux densities from FUV to NIR bands were corrected for reddening due to interstellar extinction from the MW using the method adopted by \cite{1989ApJ...345..245C}. They have derived $A(\lambda)/A(V)$ for wavelengths ranging from 0.125 to 3.5\,$\mu$m. Beyond this wavelength range, the extinction is relatively lower but significant. Hence, for accurate SED modeling, the extinction correction for the MIR bands (until 8.0\,$\mu$m) was made using the method adopted by \cite{2005ApJ...619..931I}. The UV to NIR bands help in constraining the stellar emission, whereas the MIR to mm bands constrain the dust emission from the galaxy. The 2MASS $JHK_s$ and {\sl WISE} photometry used in this work were computed in identical apertures using the custom methods described in \cite{2003AJ....125..525J} and \cite{2019ApJS..245...25J}. The radio bands are used to determine the non-thermal radio emission from the galaxy. 

\subsection{SED Modeling using CIGALE}\label{CIGALEmodeling}
The SED modeling was done using CIGALE version 0.12.1 \citep{2019A&A...622A.103B,2016A&A...585A..43C,2015A&A...576A..10C}. Written in Python, CIGALE can be used to model the UV to radio spectrum of galaxies easily and efficiently. It also estimates their physical properties such as SFR, stellar mass, attenuation, dust mass, dust luminosity, and AGN fraction. CIGALE provides the following categories of modules: SFH, Stellar Emission, Nebular Emission, Dust Attenuation, IR Re-emission, AGN Emission, Non-thermal Radio Emission, and Redshifting. Combining multiple modules, each component of the galaxy can be modeled explicitly and the physical parameters for the galaxy can be estimated. 

The SFH of the galaxies varies from quiescent phases to episodes of intense star-formation. We consider a delayed SFH where after the onset of star-formation, the SFR increases nearly linearly, reaches a peak value, and then decreases smoothly. The spectrum of the composite stellar populations is calculated using the dot product of the SFH with the grid containing the evolution of the spectrum of a single stellar population (SSP). For stellar emission, we use the \cite{2003MNRAS.344.1000B} library of SSPs and the \cite{1955ApJ...121..161S} initial mass function (IMF). The young stars emit Lyman continuum photons which ionize the surrounding gas. This energy is re-emitted in the form of a series of emission lines and a continuum, which was computed using the methods by \cite{2011MNRAS.415.2920I} and \cite{2010MNRAS.401.1325I} respectively. A major amount of the stellar emission in galaxies is attenuated by the dust grains and the energy absorbed is then re-emitted at longer wavelengths. Dust attenuation was modeled using the \cite{2000ApJ...533..682C} starburst attenuation curve, which was extended between the Lyman break and 150\,nm with the \cite{2002ApJ...574..114L} curve, adding a UV bump and a power-law. Dust emission was modeled using the \cite{2007ApJ...663..866D} model and was used to calculate the dust mass.

Along with heating the dust in their birth clouds, massive young stars also produce supernova remnants which accelerate relativistic electrons in the presence of the galactic magnetic field and produce synchrotron or non-thermal radio emission. As a consequence of this, there is a nearly linear and direct relationship between star-formation and cosmic-ray production. This forms the basis for a FIR-radio correlation (FRC; \citealp{1971A&A....15..110V,1984AJ.....89.1520R, 1985ApJ...298L...7H, 1992ARA&A..30..575C}). The processes leading to the FRC are not understood in detail, and are challenging to model \citep{2007MNRAS.379.1042V, 2010ApJ...717..196L}, but various authors have nonetheless developed empirical relations to define the FRC. The FRC is well established across many orders of magnitude in infrared as well as radio luminosities. It is independent of redshift and holds true for various galaxy types \citep{1971A&A....15..110V, 1985ApJ...298L...7H, 2001ApJ...554..803Y, 2009MNRAS.398.1573S, 2011MNRAS.410.1155B, 2015A&A...573A..45M,2018MNRAS.480.5625R}. 

CIGALE employs the FRC to relate the FIR and radio emission. The non-thermal radio emission is modeled by two parameters: the radio-IR correlation coefficient $q_\mathrm{IR}$ \citep{1985ApJ...298L...7H} and the radio power-law spectral slope $\alpha$. The values for the slope $\alpha$ were derived from the radio-frequency SEDs described in Section\,\ref{result_radio}. For the radio-IR correlation coefficient $q_\mathrm{IR}$, CIGALE provides a default value of 2.58, following \cite{1985ApJ...298L...7H} who estimated values of $q_\mathrm{IR}$ for a sample of spiral galaxies. As a result of an additional contribution from compact radio cores and radio jets/lobes, galaxies can have smaller $q_\mathrm{IR}$ values which can go as low as $\sim2$ \citep{1996ARA&A..34..749S, 2001ApJ...554..803Y}. On the other hand, extreme local starbursts are seen to have larger dispersion in $q_\mathrm{IR}$ than normal star-forming galaxies \citep{1985ApJ...298L...7H, 1991ApJ...378...65C, 2001ApJ...554..803Y}. \cite{1991ApJ...378...65C} observed high $q_\mathrm{IR}$ for extreme starbursts. The values of $q_\mathrm{IR}$ can go as high as $\sim3.0$ for starbursts and mergers \citep{2001ApJ...554..803Y}. To model the galaxies in our sample, $q_\mathrm{IR}$ was kept as a free parameter allowing for values both larger and smaller than the default value (see Table\,\ref{freepar}). 

AGN feedback has a major impact on galaxy evolution. AGN emit UV radiation which heats the dust in the torus and this thermal emission spectrum peaks usually in the MIR band. \cite{2015A&A...576A..10C} demonstrated that for the case of weak AGN, the AGN-galaxy decomposition using broad-band photometry may lead to an overestimation of the AGN fraction and hence, the AGN luminosity. To avoid this, we looked at the multi-wavelength photometry for both NGC\,7496 and NGC\,7590 in order to establish the need for an AGN component in the SED modeling process. For NGC\,7496, we also see strong emission from the core at 1300\,MHz, which supports the use of an AGN component to the model the SED of this galaxy.

\begin{figure}
\begin{center}
\includegraphics[scale=0.41]{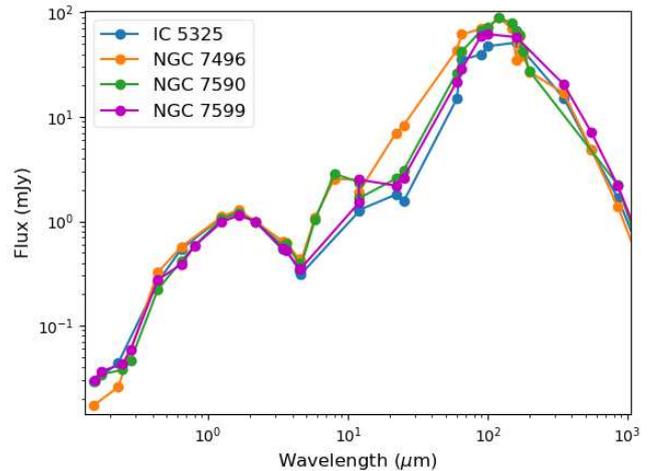}
\end{center}
\caption{The multi-wavelength SED for each of the galaxies in the 0.1-1000\,$\mu$m range, where the fluxes for each galaxy are normalized to its 2MASS K$_S$ band flux. The two Seyfert galaxies, NGC\,7496 (orange) and NGC\,7590 (green), demonstrate significantly higher fluxes in the MIR regime when compared to the two non-Seyfert galaxies, IC\,5325 (blue) and NGC\,7599 (magenta).}
\label{figure:2um_norm}
\end{figure}

Fig.\,\ref{figure:2um_norm} shows the multi-wavelength photometry for each galaxy, normalized to the 2MASS K$_S$ band flux for that galaxy. We see that the normalized optical-NIR fluxes are similar for all the galaxies but a strong distinction is apparent in the MIR regime. The two Seyfert galaxies, NGC\,7496 and NGC\,7590, show higher MIR fluxes than the other two non-AGN galaxies. This calls for the inclusion of an AGN module in the SED modeling  of both NGC\,7496 and NGC\,7590. We use the \cite{2006MNRAS.366..767F} AGN model which parameterizes the dust emission from the torus heated by the AGN.

The absorption by the intergalactic medium (IGM) is computed and redshifting of the SED model is performed after computation of all the individual modules. CIGALE searches over a grid of discrete free parameter values. The free parameters used during SED modeling are given in Table\,\ref{freepar}.

\subsection{Best-fit SEDs}\label{bestfitseds}
\begin{figure*}
\begin{center}
\includegraphics[scale=0.41]{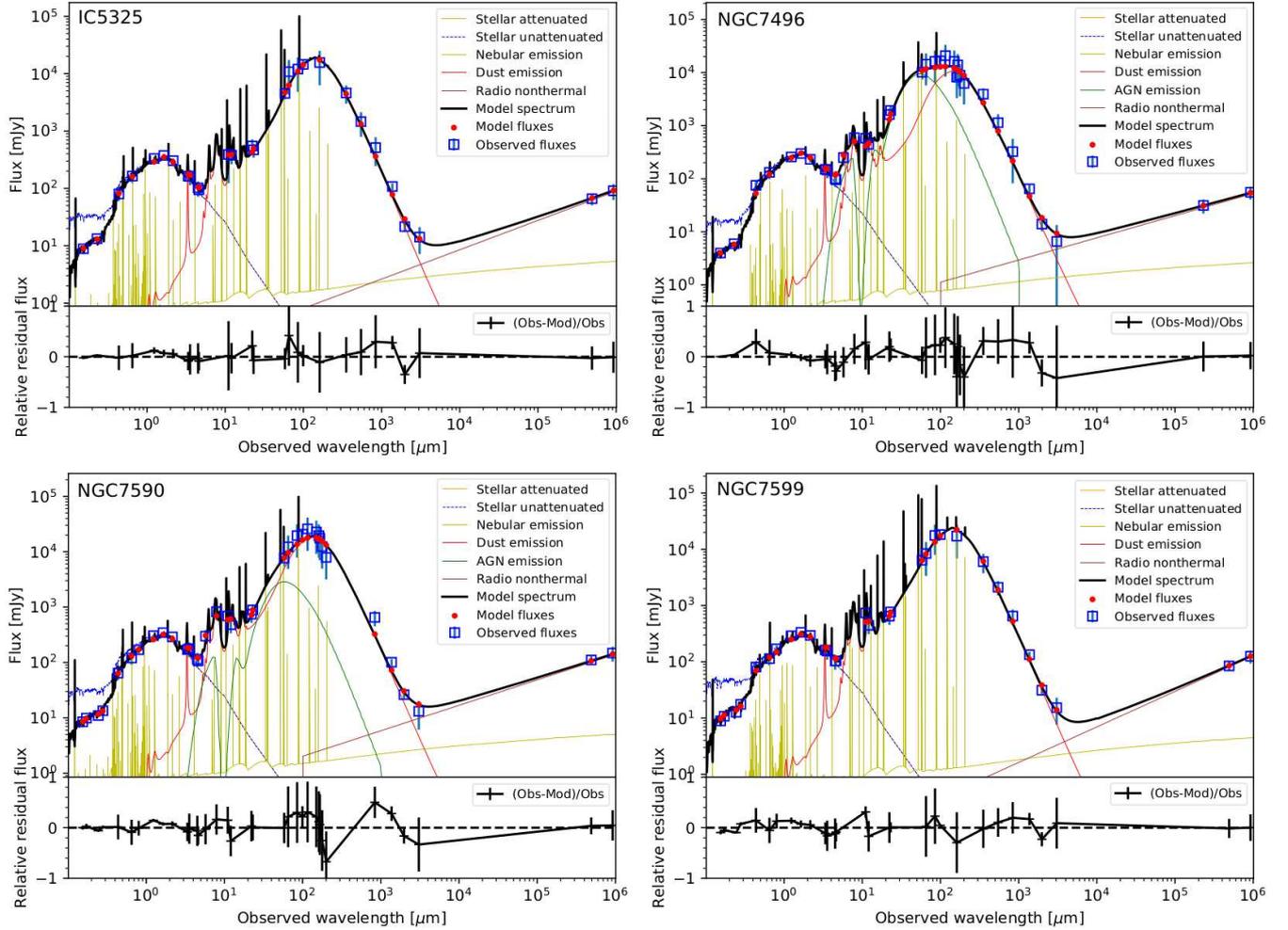}
\caption{Best-fit CIGALE SEDs for the four galaxies in our pilot sample.  The photometry is indicated with blue boxes; the models are indicated with colored lines. The models include unattenuated stellar emission (blue line), the AGN contribution (for NGC\,7495 and NGC\,7590; green line), the non-thermal radio emission, modeled with a single power-law component (maroon line), and the nebular emission (yellow-green line).  The dust absorbs stellar continuum light and re-emits at longer wavelengths, as indicated by the red line.  All model components are combined to construct a best-fit model galaxy spectrum for each galaxy, indicated in black; red dots indicate the modeled fluxes in the photometric bands.   Error bars are 3$\sigma$.}
\label{figure:bestfits}
\end{center}
\end{figure*}

The best-fit SEDs for each of our four galaxies are shown in Fig.\,\ref{figure:bestfits}, and the best-fit parameter values are listed in Table\,\ref{table:bestfitpar}. CIGALE also computes dust parameters (mass and luminosity) which are presented in Table\,\ref{table:dust}. These dust masses and luminosities have been computed for the best-fit SED models. Because CIGALE does not provide uncertainties for these quantities, we give the CIGALE Bayesian errors as indicative of the uncertainties in the best-fit parameters. The SFRs for the galaxies have been computed by CIGALE using Bayesian statistics and are listed in Table\,\ref{table:sfrcalc}. \cite{2019A&A...632A..79B} recently showed that, in order to reliably constrain galaxies' physical parameters, it is essential for SED models to closely approximate emission in the UV bands. Although the \cite{2019A&A...632A..79B} sample was at relatively high redshift, their conclusions apply generally, and are certainly germane to the analysis described here. Fortunately, for all four galaxies in our sample, the {\sl GALEX} and {\sl AstroSat} UV bands' emission is well-modeled by our CIGALE models (Fig.\,\ref{figure:bestfits}), so we can have confidence in its treatment of the energy balance and in its derivation of the basic properties for these objects. The FIR/Radio correlation coefficient for each galaxy is greater than 2.5, consistent with a lack of observed radio lobes or jets for all of them.

\begin{table*}
\begin{center}
\caption{Best-fit CIGALE model parameters for our sample galaxies.}
\label{table:bestfitpar}
\begin{tabular}{lllll}
\hline  \hline  
\textbf{Model Parameters}                                            				& \textbf{IC\,5325}		& \textbf{NGC\,7496} 	& \textbf{NGC\,7590} 	& \textbf{NGC\,7599}  	\\\hline  
\textbf{SFH}                                                                       	&                    	&                    	&                    	&                   	\\
e-folding time [{\rm Myr}] of the main stellar population model       				& 1000					& 1000                	& 1000               	& 1500                  \\
Age [{\rm Myr}] of the oldest stars in the galaxy                					& 4000					& 5000               	& 4000               	& 3500                  \\\hline  
\textbf{Stellar Emission}                                                          	&                    	&                    	&                    	&                   	\\
Initial mass function                                                              	& Salpeter           	& Salpeter           	& Salpeter           	& Salpeter          	\\
Metallicity                                  										& 0.02					& 0.02               	& 0.02               	& 0.05                  \\
Age [{\rm Myr}] of the separation between the young and the old star populations   	& 15					& 15                 	& 5                  	& 15                  	\\\hline  
\textbf{Dust Attenuation}                                                          	&                    	&                    	&                    	&                   	\\
E(B-V)*, colour excess of stellar continuum light for young population 				& 0.30					& 0.45 					& 0.25 					& 0.35         			\\
Reduction factor for the E(B-V)* of old population as compared to young one  		& 0.15 					& 0.10               	& 0.40               	& 0.10            		\\
Amplitude of the UV bump                                                           	& 0.0                  	& 0.0                  	& 0.0                  	& 0.0                 	\\
Slope delta of the power law modifying the attenuation curve               			& 0.00         			& 0.15               	& 0.00               	& 0.00              	\\\hline  
\textbf{Dust Emission}                                                             	&                    	&                    	&                    	&                  		\\
Mass fraction of PAH [\%]                               							& 3.90          		& 6.63               	& 5.95               	& 3.90         			\\
Minimum radiation field, $U_\mathrm{min}$ [Habing]                  				& 1.5          			& 1.5                	& 1.5                	& 1.0           		\\
power-law slope $dU/dM \propto U^\alpha$, $\alpha$        							& 2.3     				& 3.0                	& 2.7                	& 2.3         			\\
Fraction illuminated from $U_\mathrm{min}$ to $U_{max}$, $\gamma$      				& 0.1                   & 0.2                	& 0.4 					& 0.2                   \\\hline  
\textbf{Non-Thermal Radio Emission}                                                	&                    	&                    	&                    	&                   	\\
FIR/radio correlation coefficient                                                  	& 2.70               	& 2.64               	& 2.51               	& 2.75              	\\
Slope of non-thermal emission             											& 0.51               	& 0.42              	& 0.46             		& 0.63               	\\\hline  
\textbf{Fritz et al. (2006) : AGN Emission}                                        	&                    	&                    	&                    	&                   	\\
Ratio of the maximum to minimum radii of the dust torus          					&   -          			& 100                 	& 100                	&    -              	\\
Optical depth at 9.7 microns                                    					&    -             		& 10                 	& 10                 	&   -              		\\
$\beta$, power-law density distribution for radial component of torus 				&   -  					& -1.0               	& -1.0               	&    -              	\\
$\gamma$, power-law density distribution for polar component of torus               &   -   				& 0.0                	& 0.0                	&    -              	\\
Full opening angle of the dust torus [degree]                          				&   -            		& 140                	& 100                	&     -               	\\
Angle between equatorial axis and line of sight, $\psi$ [degree]           			& -                  	& 30.1               	& 30.1               	&     -                	\\
AGN Fraction                                                     					&   -                   & 0.6                	& 0.2            		&     -                 \\\hline  
\end{tabular}
\end{center}
\end{table*}

Below, we describe the salient features of the fits for the sources individually.
\vspace{2mm}
\\{\it \textbf{IC\,5325:}} The SFH for IC\,5325 is best-fit with the oldest stellar population to be around $4000$\,{\rm Myr} in age and with the onset of the most recent star-formation activity starting around $1000$\,{\rm Myr} ago. Stellar emission was modeled using a stellar population having a metallicity of $0.02$ and a separation age of $15$\,{\rm Myr} between the young and old stellar populations. The dust attenuation is modeled without a UV bump and exhibits a colour excess of stellar continuum light (E(B$-$V)*) for the young and old population as $0.30$ and $0.05$, respectively. The dust emission was modeled by using a power-law slope of $2.3$, a moderate PAH fraction of $3.90$, a minimum radiation field value, $U_\mathrm{min}$\,$=$\,$1.5$ and $90$\% of dust is exposed to the radiation between $U_\mathrm{min}$ and $U_{max}$. The slope of the synchrotron emission is $\sim$\,$0.5$, and the FIR/radio correlation coefficient, $q_\mathrm{IR}$\,$=$\,$2.70$. The dust mass is estimated to be $3.75\times10^7$\,M$_\odot$.  
\vspace{2mm}
\\{\it \textbf{NGC\,7496:}} NGC\,7496 is best modeled with an older stellar population of $5000$\,{\rm Myr} age with the most recent burst of star-formation taking place around $1000$\,{\rm Myr} ago. The stellar population is modeled with solar metallicity and a separation of $15$\,{\rm Myr} between the young and old stellar population. The dust attenuation is modeled without a UV bump and exhibits a colour excess of stellar continuum light (E(B$-$V)*) for young and old population as $0.45$ and $0.05$, respectively. The slope delta of the power law modifying the attenuation curve is $0.15$. The dust emission is best modeled by using a power-law slope of $3.0$, a very high PAH fraction of $6.63$ and a minimum radiation field value, $U_\mathrm{min}$\,$=$\,$1.5$, with $80$\% of dust being exposed to the radiation between $U_\mathrm{min}$ and $U_{max}$. The synchrotron emission was modeled with a FIR/radio correlation coefficient $q_\mathrm{IR}$\,$=$\,$2.64$, and slope of the power-law, $\alpha$\,$=$\,$0.42$. Emission from the dusty torus of the AGN is modeled with a decreasing dust density distribution ($\propto$\,$1/r$) along the radial direction. The galaxy exhibits a high AGN fraction ($0.6$) which implies higher luminosity illuminating the dusty torus. The dust mass for NGC\,7496 is estimated to be $2.70\times10^7$\,M$_\odot$, indicating that it is a dusty galaxy. Some excess emission is seen in the submm/mm region, which hints the presence of very cold dust. Thus, the calculated dust mass is a lower limit to the total galaxy\rq s dust mass. 
\vspace{2mm}
\\{\it \textbf{NGC\,7590:}} The SFH for NGC\,7590 is best modeled with an old stellar population of $4000$\,{\rm Myr} age and with the most recent star-formation activity starting $1000$\,{\rm Myr} ago. The stellar population is modeled having the solar metallicity with the difference in the ages of young and old stellar populations to be $5$\,{\rm Myr}. The attenuation curve is modeled without a UV bump and with a colour excess of stellar light (E(B$-$V)*) for young and old population as $0.25$ and $0.10$, respectively. The dust emission is modeled as a power-law having a slope of $2.7$, a high PAH fraction of $5.95$, a minimum radiation field value $U_\mathrm{min}$\,$=$\,$1.5$, and with $60$\% of dust being exposed to radiation between $U_\mathrm{min}$ and $U_{max}$. The slope of the power-law non-thermal radio emission is taken to be $\sim$\,$0.5$, as obtained from the radio continuum spectra of the galaxy and the corresponding best-fit FIR/radio correlation coefficient is $q_\mathrm{IR}$\,$=$\,$2.51$. The emission from the dusty torus of the AGN was modeled using a decreasing dust density distribution ($\propto$\,$1/r$) along the radial direction. This galaxy exhibits a low AGN fraction ($0.2$) and a larger torus than NGC\,7496, but has a lower dust luminosity which could be due to the insufficient heating of dust. The dust mass for the galaxy is estimated to be $3.57\times10^7$\,M$_\odot$. From the SED, it is clearly evident that the {\sl Planck} HFI data point are not well-fit. This is a piece of evidence for the presence of a very cold dust component. Hence, the calculated dust mass is a lower limit to the total galaxy\rq s dust mass.
\vspace{2mm}
\\{\it \textbf{NGC\,7599:}} The SED for NGC\,7599 is best modeled having a SFH with the oldest stars being $3500$\,{\rm Myr} in age and the last star-formation activity starting $1500$\,{\rm Myr} ago. The stellar population was best modeled with the young and old stellar population having a metallicity of $0.05$ and an age separation of $15$\,{\rm Myr}. Using the \cite{2000ApJ...533..682C} dust attenuation law without a UV bump, the best-fit corresponds to colour excess (E(B$-$V)*) values of $0.35$ and $0.04$ for young and old stellar populations, respectively. The dust emission was modeled using a power-law slope of $2.3$, a moderate  PAH fraction of $3.90$, a minimum radiation field value $U_\mathrm{min}$\,$=$\,$1.0$, and $80$\% of dust is exposed to the radiation between $U_\mathrm{min}$ and $U_{max}$. For the synchrotron emission, the best-fit corresponds to a slope of $\sim$\,$0.6$ and the FIR/radio correlation coefficient, $q_\mathrm{IR}$\,$=$\,$2.75$. The dust mass for the galaxy estimated from the best-fit is $7.17\times10^7$\,M$_\odot$, which suggests that the galaxy is a very dusty galaxy and has the highest dust content in our sample of galaxies.   

\begin{figure*}
\begin{center}
\includegraphics[scale=0.28]{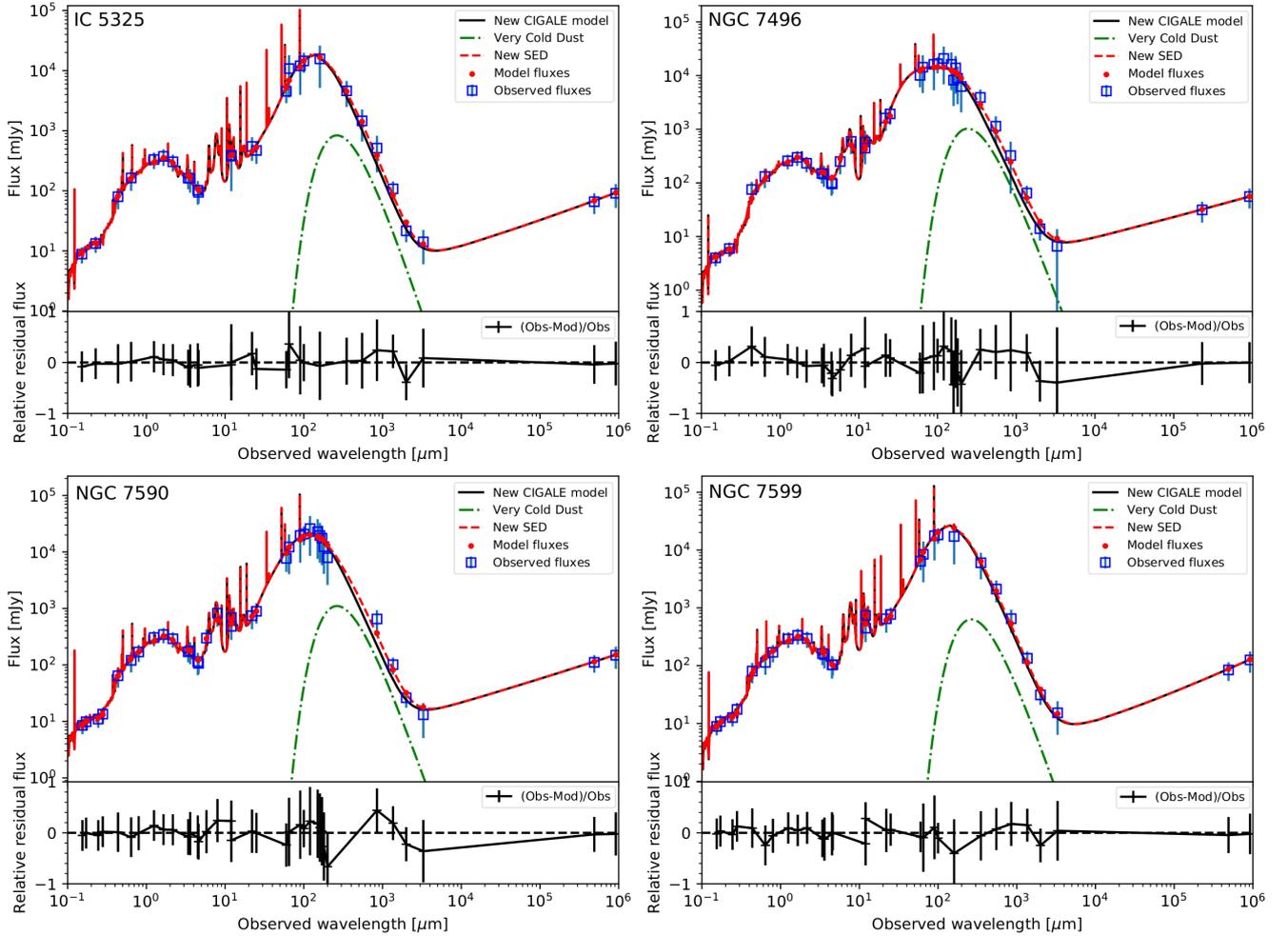}
\end{center}
\caption{The new SEDs for each of the galaxies obtained (dashed red line) after adding the modified CIGALE model (solid black line) and the very cold dust component (dot-dashed green line).}
\label{figure:vcd}
\end{figure*}

\section{VERY COLD DUST}\label{vcd}
As is evident from Fig.\,\ref{figure:bestfits}, NGC\,7496 and NGC\,7590 show evidence of excess emission in the submm/mm regime, relative to the best-fit CIGALE SED models. This argues for the presence of very cold dust in the galaxies. Such a submm excess is also seen in various low-metallicity galaxies \citep{1995A&A...295..317C,1998A&A...331L...9K,2003A&A...407..159G,2005A&A...434..867G,2006ApJ...646..929M}. \cite{1996ASSL..209...21D} first suggested that dust grains having higher IR and FIR emissivities can achieve very low temperatures. The very cold dust grains are embedded deeply in the clumpy and dense molecular clouds and they are illuminated by the FIR emission from the \lq classical grains\rq\space \citep{2003A&A...407..159G, 2004A&A...414..475D}. Another possible explanation for these dust grains could be due to  the dust grains having some unusual optical properties such as fractal or porous grains having high submm/mm emissivity \citep{1995ApJ...451..188R, 2004A&A...414..475D}. But these explanations for the very low temperature of dust grains are inconsistent with the fact that observations show very cold dust is distributed over galactic scales. 

\begin{table}
\caption{Calculated parameters for dust from the best-fit CIGALE SEDs.}
\label{table:dust}
\begin{center}
\begin{tabular}{l  c c c}
\hline
\hline
\textbf{Galaxy} & \textbf{Dust Mass}			& \textbf{Dust Luminosity}	& \textbf{AGN Torus} 	\\
 				&  								& 							& \textbf{Luminosity} 	\\
				& $10^{7}$ M$_\odot$ 			& $10^{10}$\,L$_\odot$ 		& $10^{10}$\,L$_\odot$ 	\\\hline
IC\,5325  		& $3.75 \pm 0.26$ 				& $1.09\pm 0.05$ 			&	-					\\
NGC\,7496 		& $2.70 \pm 0.30$ 				& $0.79\pm 0.04$ 			& $1.20\pm 0.06$ 		\\
NGC\,7590 		& $3.57 \pm 0.38$ 				& $1.32\pm 0.07$ 			& $0.35\pm 0.02$ 		\\
NGC\,7599 		& $7.17 \pm 0.61$ 				& $1.74\pm 0.09$ 			&	-					\\\hline
\end{tabular}
\end{center}
\end{table}

To estimate the very cold dust content in the galaxies in our sample, we added an additional very cold thermal dust component independent to the CIGALE models. The luminosity of the very cold dust component is calculated using the relation given by \cite{2011A&A...532A..56G}:
\begin{equation}
\mathrm{L^\mathrm{CD}}_\nu(\lambda) =  \mathrm{M}_\mathrm{CD}3\pi \Big(\frac{1}{\rho}\Big) \Big(\frac{Q_0}{a}\Big) \lambda^\beta_0 [\lambda^{-\beta} B_\nu(\lambda,\mathrm{T}_\mathrm{CD})]
\end{equation}
where $\mathrm{M}_\mathrm{CD}$ is the mass of the very cold dust, $B_\nu$ is the Planck function, $\mathrm{T}_\mathrm{CD}$ is the very cold dust temperature, $\beta$\,$=$\,$2$ is the emissivity coefficient, $\lambda_0$\,$=$\,$100$\,$\mu$m is the reference wavelength, $\rho$\,$=$\,$2.5 \times 10^3$\,kg\,m$^{-3}$ is the mass density of the grains, and $Q_0/a$\,$=$\,$150$\,cm$^{-1}$ at $\lambda_0$ is the absorption coefficient with $a$ as the radius of the grain. We first updated a few parameters in CIGALE to generate the CIGALE SEDs, which were then read using a Python code, and a very cold dust component added to constrain submm/mm observations and get the new SEDs which are shown in Fig.\,\ref{figure:vcd}. Based on our initial SED modeling, the global photometry for our mm-selected galaxies is consistent with total dust masses as given in Table\,\ref{table:dust}. However, SEDs do not rule out additional very cold dust components, the masses of which are likewise given in Table\,\ref{table:vcdtab} for the models that include very cold dust components.  Based on the goodness of fit statistics, the cold dust components did not improve the fits significantly for IC\,5325 and NGC\,7599.   However, for NGC\,7496 and NGC\,7590 the additional very cold dust component does improve the best-fit SEDs marginally, but the very cold dust does not dominate the total dust mass.

\begin{table*}
\begin{center}
\caption{Best-fit very cold dust (VCD) temperatures and masses from the refined modeling.}
\label{table:vcdtab}
\begin{tabular}{lcccccc}
\hline
\textbf{Galaxy} & \textbf{Dust Mass} 	& \textbf{$\boldsymbol{\mathrm{T}_\mathrm{VCD}^*}$} 	& \textbf{VCD Mass} 		& \textbf{Without VCD}		& \textbf{With VCD}		& $\boldsymbol{\triangle \chi^2}$\\
                & (10$^7$\,M$_\odot$)	& (K) 				  					& (10$^7$\,M$_\odot$)   	& $\chi^2/$DOF					& $\chi^2/$DOF			&					\\\hline \vspace{1mm}
IC\,5325        & $2.84 \pm 0.17$       & $11^{+2.05}_{-3.22}$  				& $1.20^{+0.32}_{-0.56}$	& $28.35/17$					& $27.54/15$				& $0.81$				\\\vspace{1mm}
NGC\,7496       & $1.63 \pm 0.08$       & $12^{+0.99}_{-1.16}$  				& $1.20^{+0.29}_{-0.30}$	& $61.20/18$					& $55.42/16$				& $5.78$			\\\vspace{1mm}
NGC\,7590       & $2.15 \pm 0.11$       & $11^{+0.79}_{-1.90}$  				& $1.78^{+0.67}_{-0.36}$    & $55.65/19$					& $49.35/17$				& $6.30$			\\\vspace{1mm}
NGC\,7599       & $5.67 \pm 0.28$       & $11^{+1.88}_{-2.74}$  				& $1.14^{+0.58}_{-0.58}$    & $34.20/20$					& $33.90/18$				& $0.30$			\\\hline
\end{tabular}
\end{center}
\begin{flushleft}
$^*\mathrm{T}_\mathrm{VCD}$ is the temperature of the very cold dust component.
\end{flushleft}
\end{table*}

Without the constraining submm/mm observations, the total dust mass could be underestimated by an order of magnitude \citep{2005A&A...434..867G, 2010A&A...518L..89G}. As dust grains significantly affect the molecular formation rate, cooling of the ISM as well as the SFRs in a galaxy \citep{1979ApJS...41..555H, 2013EP&S...65..213A}, an inaccurate estimate of dust mass in a galaxy would lead to an erroneous interpretation of the evolution of galaxies. \cite{1990ApJ...359...42D} suggested that irrespective of the ISM\rq s phase, most of the dust grains in spiral galaxies have a temperature $<$\,15\,K. These very cold dust grains comprise 40 to 70\% of the total dust mass in the galaxy. In our sample of four galaxies, the very cold dust comprises no more than 15 to 45\% of the total dust mass in the galaxies. Two galaxies, IC\,5325 and NGC\,7599, do not have significant very cold dust masses. On the other hand, very cold dust accounts for nearly half, $40-45$\% of the total dust content in NGC\,7496 and NGC\,7590, respectively. This is consistent with the values derived by \cite{1990ApJ...359...42D} for a similar sample of low-redshift spiral galaxies.

\section{SUMMARY}\label{summary}
We examined some southern star-forming galaxies (IC\,5325, NGC\,7496, NGC\,7590, and NGC\,7599) in bands ranging from UV to radio. Using the GMRT observations, the radio fluxes were calculated for all the galaxies and the radio imaging of the galaxies was performed. The radio flux densities were also used to calculate the SFRs for each of the galaxies which indicated that all the galaxies in the sample are moderate star-forming galaxies just like the MW. The radio fluxes from GMRT as well as other radio telescopes were used to estimate the radio spectral indices for the galaxies. The UVIT imaging of two of the galaxies (NGC\,7590 and NGC\,7599) in the sample was used to identify the young population in these galaxies. In NGC\,7599, we also report the presence of an anamolous star-forming region which could be a signature of triggered star-formation due to the interaction with the neighboring galaxy, NGC\,7590  (Section\,\ref{trigsf}). SED fitting using CIGALE was used to estimate dust masses for each galaxy. We found that all four galaxies in the sample have significant warmer dust components (Table\,\ref{table:dust}). To estimate the sample galaxies' very cold dust content, we introduced a simple additional very cold dust component into our modeling framework and quantified the maximum very cold dust masses that these galaxies could contain (Table\,\ref{table:vcdtab}). Very cold dust mass does not dominate the dust mass budgets of galaxies in our sample. Among the four, NGC\,7496 and NGC\,7590 show significant improvement in the SED fits when a very cold dust component is included. The dust masses are highly sensitive to the SED fits in the radio regime. This study, thus, highlights the importance of radio data for precise estimation of star-forming galaxies' dust content. We encourage more such studies, spanning large wavelength ranges with bigger galaxy samples for accurate estimation of cold dust content in star-forming galaxies so that its relationship to star-formation activity can be better understood.

\section*{ACKNOWLEDGEMENTS}
\small We thank the staff of GMRT, National Centre for Radio Astrophysics (NCRA) - Tata Institute of Fundamental Research (TIFR), who made the radio observations possible. This publication also uses data from the {\sl AstroSat} mission of the Indian Space Research Organisation (ISRO), archived at the Indian Space Science Data Centre (ISSDC). This work also uses data from the SPT survey. SPT is supported by the National Science Foundation through grants PLR-1248097 and OPP-1852617. Partial support is also provided by the NSF Physics Frontier Center grant PHY-1125897 to the Kavli Institute of Cosmological Physics at the University of Chicago, the Kavli Foundation and the Gordon and Betty Moore Foundation grant GBMF 947.

This work is based in part on observations made with the {\sl Spitzer Space Telescope}, which is operated by the Jet Propulsion Laboratory, California Institute of Technology under a contract with NASA. This publication makes use of data products from the {\sl Wide-field Infrared Survey Explorer}, which is a joint project of the University of California, Los Angeles, and the Jet Propulsion Laboratory/California Institute of Technology, funded by NASA. This research uses data products of the Two Micron All Sky Survey, which is a joint project of the University of Massachusetts and the Infrared Processing and Analysis Center/California Institute of Technology, funded by NASA and the NSF. This work also uses data from {\sl GALEX} and SUMSS surveys. {\sl GALEX} (Galaxy Evolution Explorer) is a NASA Small Explorer, launched in April 2003. We gratefully acknowledge NASA\rq s support for construction, operation, and science analysis for the {\sl GALEX} mission, developed in cooperation with the Centre National d\rq Etudes Spatiales of France and the Korean Ministry of Science and Technology.  We thank the Molonglo Observatory staff who are responsible for the smooth operation of the Molonglo Observatory Synthesis Telescope (MOST) and the day-to-day observing program of SUMSS. This work uses observations obtained with Planck (http://www.esa.int/Planck), an ESA science mission with instruments and contributions directly funded by ESA Member States, NASA, and Canada. This research has used  data products from {\sl AKARI}, a JAXA project with the participation of ESA. This research has also made use of the NASA/IPAC Extragalactic Database (NED) which is operated by the Jet Propulsion Laboratory, California Institute of Technology, under contract with NASA. 

The authors gratefully thank the Referee and the Editors for their constructive comments and suggestions which helped improve the quality of this paper. We thank Antony Stark for his valuable comments on the manuscript. SS thanks GH, SAG; DD, PDMSA and Director, URSC for encouragement and continuous support to carry out this research.

\normalsize

\section*{DATA AVAILABILTY}
\small
The data underlying this article are available in the article and the references therein. 
\normalsize



\bibliographystyle{mnras}
\bibliography{references} 

\begin{thebibliography}{}
\makeatletter
\relax
\def\mn@urlcharsother{\let\do\@makeother \do\$\do\&\do\#\do\^\do\_\do\%\do\~}
\def\mn@doi{\begingroup\mn@urlcharsother \@ifnextchar [ {\mn@doi@}
  {\mn@doi@[]}}
\def\mn@doi@[#1]#2{\def\@tempa{#1}\ifx\@tempa\@empty \href
  {http://dx.doi.org/#2} {doi:#2}\else \href {http://dx.doi.org/#2} {#1}\fi
  \endgroup}
\def\mn@eprint#1#2{\mn@eprint@#1:#2::\@nil}
\def\mn@eprint@arXiv#1{\href {http://arxiv.org/abs/#1} {{\tt arXiv:#1}}}
\def\mn@eprint@dblp#1{\href {http://dblp.uni-trier.de/rec/bibtex/#1.xml}
  {dblp:#1}}
\def\mn@eprint@#1:#2:#3:#4\@nil{\def\@tempa {#1}\def\@tempb {#2}\def\@tempc
  {#3}\ifx \@tempc \@empty \let \@tempc \@tempb \let \@tempb \@tempa \fi \ifx
  \@tempb \@empty \def\@tempb {arXiv}\fi \@ifundefined
  {mn@eprint@\@tempb}{\@tempb:\@tempc}{\expandafter \expandafter \csname
  mn@eprint@\@tempb\endcsname \expandafter{\@tempc}}}

\bibitem[\protect\citeauthoryear{{Agrawal}}{{Agrawal}}{2004}]{2004PThPS.155..305A}
{Agrawal} P.~C.,  2004, \mn@doi [Progress of Theoretical Physics Supplement]
  {10.1143/PTPS.155.305}, \href
  {http://adsabs.harvard.edu/abs/2004PThPS.155..305A} {155, 305}

\bibitem[\protect\citeauthoryear{{Asano}, {Takeuchi}, {Hirashita}  \&
  {Inoue}}{{Asano} et~al.}{2013}]{2013EP&S...65..213A}
{Asano} R.~S.,  {Takeuchi} T.~T.,  {Hirashita} H.,   {Inoue} A.~K.,  2013,
  \mn@doi [Earth, Planets, and Space] {10.5047/eps.2012.04.014}, \href
  {https://ui.adsabs.harvard.edu/abs/2013EP&S...65..213A} {65, 213}

\bibitem[\protect\citeauthoryear{{Bendo} et~al.,}{{Bendo}
  et~al.}{2012}]{2012MNRAS.419.1833B}
{Bendo} G.~J.,  et~al., 2012, \mn@doi [\mnras]
  {10.1111/j.1365-2966.2011.19735.x}, \href
  {https://ui.adsabs.harvard.edu/abs/2012MNRAS.419.1833B} {419, 1833}

\bibitem[\protect\citeauthoryear{{Boquien}, {Burgarella}, {Roehlly}, {Buat},
  {Ciesla}, {Corre}, {Inoue}  \& {Salas}}{{Boquien}
  et~al.}{2019}]{2019A&A...622A.103B}
{Boquien} M.,  {Burgarella} D.,  {Roehlly} Y.,  {Buat} V.,  {Ciesla} L.,
  {Corre} D.,  {Inoue} A.~K.,   {Salas} H.,  2019, \mn@doi [\aap]
  {10.1051/0004-6361/201834156}, \href
  {https://ui.adsabs.harvard.edu/abs/2019A&A...622A.103B} {622, A103}

\bibitem[\protect\citeauthoryear{{Bot}, {Ysard}, {Paradis}, {Bernard},
  {Lagache}, {Israel}  \& {Wall}}{{Bot} et~al.}{2010}]{2010A&A...523A..20B}
{Bot} C.,  {Ysard} N.,  {Paradis} D.,  {Bernard} J.~P.,  {Lagache} G.,
  {Israel} F.~P.,   {Wall} W.~F.,  2010, \mn@doi [\aap]
  {10.1051/0004-6361/201014986}, \href
  {http://adsabs.harvard.edu/abs/2010A%26A...523A..20B} {523, A20}

\bibitem[\protect\citeauthoryear{{Bourne}, {Dunne}, {Ivison}, {Maddox},
  {Dickinson}  \& {Frayer}}{{Bourne} et~al.}{2011}]{2011MNRAS.410.1155B}
{Bourne} N.,  {Dunne} L.,  {Ivison} R.~J.,  {Maddox} S.~J.,  {Dickinson} M.,
  {Frayer} D.~T.,  2011, \mn@doi [\mnras] {10.1111/j.1365-2966.2010.17517.x},
  \href {https://ui.adsabs.harvard.edu/abs/2011MNRAS.410.1155B} {410, 1155}

\bibitem[\protect\citeauthoryear{{Bruzual} \& {Charlot}}{{Bruzual} \&
  {Charlot}}{2003}]{2003MNRAS.344.1000B}
{Bruzual} G.,  {Charlot} S.,  2003, \mn@doi [\mnras]
  {10.1046/j.1365-8711.2003.06897.x}, \href
  {http://adsabs.harvard.edu/abs/2003MNRAS.344.1000B} {344, 1000}

\bibitem[\protect\citeauthoryear{{Buat} et~al.,}{{Buat}
  et~al.}{2010}]{2010MNRAS.409L...1B}
{Buat} V.,  et~al., 2010, \mn@doi [\mnras] {10.1111/j.1745-3933.2010.00916.x},
  \href {http://adsabs.harvard.edu/abs/2010MNRAS.409L...1B} {409, L1}

\bibitem[\protect\citeauthoryear{{Buat}, {Ciesla}, {Boquien}, {Ma{\l}ek}  \&
  {Burgarella}}{{Buat} et~al.}{2019}]{2019A&A...632A..79B}
{Buat} V.,  {Ciesla} L.,  {Boquien} M.,  {Ma{\l}ek} K.,   {Burgarella} D.,
  2019, \mn@doi [\aap] {10.1051/0004-6361/201936643}, \href
  {https://ui.adsabs.harvard.edu/abs/2019A&A...632A..79B} {632, A79}

\bibitem[\protect\citeauthoryear{{Calzetti}, {Armus}, {Bohlin}, {Kinney},
  {Koornneef}  \& {Storchi-Bergmann}}{{Calzetti}
  et~al.}{2000}]{2000ApJ...533..682C}
{Calzetti} D.,  {Armus} L.,  {Bohlin} R.~C.,  {Kinney} A.~L.,  {Koornneef} J.,
   {Storchi-Bergmann} T.,  2000, \mn@doi [\apj] {10.1086/308692}, \href
  {http://adsabs.harvard.edu/abs/2000ApJ...533..682C} {533, 682}

\bibitem[\protect\citeauthoryear{{Cardelli}, {Clayton}  \& {Mathis}}{{Cardelli}
  et~al.}{1989}]{1989ApJ...345..245C}
{Cardelli} J.~A.,  {Clayton} G.~C.,   {Mathis} J.~S.,  1989, \mn@doi [\apj]
  {10.1086/167900}, \href {http://adsabs.harvard.edu/abs/1989ApJ...345..245C}
  {345, 245}

\bibitem[\protect\citeauthoryear{{Carlstrom} et~al.,}{{Carlstrom}
  et~al.}{2011}]{2011PASP..123..568C}
{Carlstrom} J.~E.,  et~al., 2011, \mn@doi [\pasp] {10.1086/659879}, \href
  {http://adsabs.harvard.edu/abs/2011PASP..123..568C} {123, 568}

\bibitem[\protect\citeauthoryear{{Casey}, {Narayanan}  \& {Cooray}}{{Casey}
  et~al.}{2014}]{2014PhR...541...45C}
{Casey} C.~M.,  {Narayanan} D.,   {Cooray} A.,  2014, \mn@doi [\physrep]
  {10.1016/j.physrep.2014.02.009}, \href
  {http://adsabs.harvard.edu/abs/2014PhR...541...45C} {541, 45}

\bibitem[\protect\citeauthoryear{{Chini}, {Kruegel}, {Lemke}  \&
  {Ward-Thompson}}{{Chini} et~al.}{1995}]{1995A&A...295..317C}
{Chini} R.,  {Kruegel} E.,  {Lemke} R.,   {Ward-Thompson} D.,  1995, \aap,
  \href {https://ui.adsabs.harvard.edu/abs/1995A&A...295..317C} {295, 317}

\bibitem[\protect\citeauthoryear{{Chomiuk} \& {Povich}}{{Chomiuk} \&
  {Povich}}{2011}]{2011AJ....142..197C}
{Chomiuk} L.,  {Povich} M.~S.,  2011, \mn@doi [\aj]
  {10.1088/0004-6256/142/6/197}, \href
  {http://adsabs.harvard.edu/abs/2011AJ....142..197C} {142, 197}

\bibitem[\protect\citeauthoryear{{Ciesla} et~al.,}{{Ciesla}
  et~al.}{2014}]{2014A&A...565A.128C}
{Ciesla} L.,  et~al., 2014, \mn@doi [\aap] {10.1051/0004-6361/201323248}, \href
  {http://adsabs.harvard.edu/abs/2014A%26A...565A.128C} {565, A128}

\bibitem[\protect\citeauthoryear{{Ciesla} et~al.,}{{Ciesla}
  et~al.}{2015}]{2015A&A...576A..10C}
{Ciesla} L.,  et~al., 2015, \mn@doi [\aap] {10.1051/0004-6361/201425252}, \href
  {https://ui.adsabs.harvard.edu/abs/2015A&A...576A..10C} {576, A10}

\bibitem[\protect\citeauthoryear{{Ciesla} et~al.,}{{Ciesla}
  et~al.}{2016}]{2016A&A...585A..43C}
{Ciesla} L.,  et~al., 2016, \mn@doi [\aap] {10.1051/0004-6361/201527107}, \href
  {https://ui.adsabs.harvard.edu/abs/2016A&A...585A..43C} {585, A43}

\bibitem[\protect\citeauthoryear{{Clemens} et~al.,}{{Clemens}
  et~al.}{2013}]{2013MNRAS.433..695C}
{Clemens} M.~S.,  et~al., 2013, \mn@doi [\mnras] {10.1093/mnras/stt760}, \href
  {http://adsabs.harvard.edu/abs/2013MNRAS.433..695C} {433, 695}

\bibitem[\protect\citeauthoryear{{Cluver}, {Jarrett}, {Dale}, {Smith}, {August}
   \& {Brown}}{{Cluver} et~al.}{2017}]{2017ApJ...850...68C}
{Cluver} M.~E.,  {Jarrett} T.~H.,  {Dale} D.~A.,  {Smith} J. D.~T.,  {August}
  T.,   {Brown} M.~J.~I.,  2017, \mn@doi [\apj] {10.3847/1538-4357/aa92c7},
  \href {https://ui.adsabs.harvard.edu/abs/2017ApJ...850...68C} {850, 68}

\bibitem[\protect\citeauthoryear{{Condon}}{{Condon}}{1992}]{1992ARA&A..30..575C}
{Condon} J.~J.,  1992, \mn@doi [\araa] {10.1146/annurev.aa.30.090192.003043},
  \href {http://adsabs.harvard.edu/abs/1992ARA%26A..30..575C} {30, 575}

\bibitem[\protect\citeauthoryear{{Condon}, {Huang}, {Yin}  \& {Thuan}}{{Condon}
  et~al.}{1991}]{1991ApJ...378...65C}
{Condon} J.~J.,  {Huang} Z.~P.,  {Yin} Q.~F.,   {Thuan} T.~X.,  1991, \mn@doi
  [\apj] {10.1086/170407}, \href
  {https://ui.adsabs.harvard.edu/abs/1991ApJ...378...65C} {378, 65}

\bibitem[\protect\citeauthoryear{{Cox}, {Kruegel}  \& {Mezger}}{{Cox}
  et~al.}{1986}]{1986A&A...155..380C}
{Cox} P.,  {Kruegel} E.,   {Mezger} P.~G.,  1986, \aap, \href
  {http://adsabs.harvard.edu/abs/1986A%26A...155..380C} {155, 380}

\bibitem[\protect\citeauthoryear{{Dale} et~al.,}{{Dale}
  et~al.}{2012}]{2012ApJ...745...95D}
{Dale} D.~A.,  et~al., 2012, \mn@doi [\apj] {10.1088/0004-637X/745/1/95}, \href
  {http://adsabs.harvard.edu/abs/2012ApJ...745...95D} {745, 95}

\bibitem[\protect\citeauthoryear{{Devereux} \& {Young}}{{Devereux} \&
  {Young}}{1990}]{1990ApJ...359...42D}
{Devereux} N.~A.,  {Young} J.~S.,  1990, \mn@doi [\apj] {10.1086/169031}, \href
  {https://ui.adsabs.harvard.edu/abs/1990ApJ...359...42D} {359, 42}

\bibitem[\protect\citeauthoryear{{Devriendt}, {Guiderdoni}  \&
  {Sadat}}{{Devriendt} et~al.}{1999}]{1999A&A...350..381D}
{Devriendt} J.~E.~G.,  {Guiderdoni} B.,   {Sadat} R.,  1999, \aap, \href
  {https://ui.adsabs.harvard.edu/abs/1999A&A...350..381D} {350, 381}

\bibitem[\protect\citeauthoryear{{Disney}}{{Disney}}{1996}]{1996ASSL..209...21D}
{Disney} M.,  1996, in {Block} D.~L.,  {Greenberg} J.~M.,  eds,  Astrophysics
  and Space Science Library Vol. 209, New Extragalactic Perspectives in the New
  South Africa. p.~21, \mn@doi{10.1007/978-94-009-0335-7_2}

\bibitem[\protect\citeauthoryear{{Draine} \& {Lazarian}}{{Draine} \&
  {Lazarian}}{1998}]{1998ApJ...508..157D}
{Draine} B.~T.,  {Lazarian} A.,  1998, \mn@doi [\apj] {10.1086/306387}, \href
  {https://ui.adsabs.harvard.edu/abs/1998ApJ...508..157D} {508, 157}

\bibitem[\protect\citeauthoryear{{Draine} et~al.,}{{Draine}
  et~al.}{2007}]{2007ApJ...663..866D}
{Draine} B.~T.,  et~al., 2007, \mn@doi [\apj] {10.1086/518306}, \href
  {http://adsabs.harvard.edu/abs/2007ApJ...663..866D} {663, 866}

\bibitem[\protect\citeauthoryear{{Dumke}, {Krause}  \& {Wielebinski}}{{Dumke}
  et~al.}{2004}]{2004A&A...414..475D}
{Dumke} M.,  {Krause} M.,   {Wielebinski} R.,  2004, \mn@doi [\aap]
  {10.1051/0004-6361:20031636}, \href
  {https://ui.adsabs.harvard.edu/abs/2004A&A...414..475D} {414, 475}

\bibitem[\protect\citeauthoryear{{Everett} et~al.,}{{Everett}
  et~al.}{2020}]{2020arXiv200303431E}
{Everett} W.~B.,  et~al., 2020, arXiv e-prints, \href
  {https://ui.adsabs.harvard.edu/abs/2020arXiv200303431E} {p. arXiv:2003.03431}

\bibitem[\protect\citeauthoryear{{Farnes}, {Green}  \& {Kantharia}}{{Farnes}
  et~al.}{2014}]{2014MNRAS.437.3236F}
{Farnes} J.~S.,  {Green} D.~A.,   {Kantharia} N.~G.,  2014, \mn@doi [\mnras]
  {10.1093/mnras/stt2118}, \href
  {https://ui.adsabs.harvard.edu/#abs/2014MNRAS.437.3236F} {437, 3236}

\bibitem[\protect\citeauthoryear{{Fazio} et~al.,}{{Fazio}
  et~al.}{2004}]{2004ApJS..154...10F}
{Fazio} G.~G.,  et~al., 2004, \mn@doi [\apjs] {10.1086/422843}, \href
  {https://ui.adsabs.harvard.edu/abs/2004ApJS..154...10F} {154, 10}

\bibitem[\protect\citeauthoryear{{Freeland}, {Stilp}  \& {Wilcots}}{{Freeland}
  et~al.}{2009}]{2009AJ....138..295F}
{Freeland} E.,  {Stilp} A.,   {Wilcots} E.,  2009, \mn@doi [\aj]
  {10.1088/0004-6256/138/1/295}, \href
  {https://ui.adsabs.harvard.edu/#abs/2009AJ....138..295F} {138, 295}

\bibitem[\protect\citeauthoryear{{Fritz}, {Franceschini}  \&
  {Hatziminaoglou}}{{Fritz} et~al.}{2006}]{2006MNRAS.366..767F}
{Fritz} J.,  {Franceschini} A.,   {Hatziminaoglou} E.,  2006, \mn@doi [\mnras]
  {10.1111/j.1365-2966.2006.09866.x}, \href
  {http://adsabs.harvard.edu/abs/2006MNRAS.366..767F} {366, 767}

\bibitem[\protect\citeauthoryear{{Galametz} et~al.,}{{Galametz}
  et~al.}{2010}]{2010A&A...518L..55G}
{Galametz} M.,  et~al., 2010, \mn@doi [\aap] {10.1051/0004-6361/201014604},
  \href {https://ui.adsabs.harvard.edu/abs/2010A&A...518L..55G} {518, L55}

\bibitem[\protect\citeauthoryear{{Galametz}, {Madden}, {Galliano}, {Hony},
  {Bendo}  \& {Sauvage}}{{Galametz} et~al.}{2011}]{2011A&A...532A..56G}
{Galametz} M.,  {Madden} S.~C.,  {Galliano} F.,  {Hony} S.,  {Bendo} G.~J.,
  {Sauvage} M.,  2011, \mn@doi [\aap] {10.1051/0004-6361/201014904}, \href
  {http://adsabs.harvard.edu/abs/2011A%26A...532A..56G} {532, A56}

\bibitem[\protect\citeauthoryear{{Galametz} et~al.,}{{Galametz}
  et~al.}{2014}]{2014MNRAS.439.2542G}
{Galametz} M.,  et~al., 2014, \mn@doi [\mnras] {10.1093/mnras/stu113}, \href
  {http://adsabs.harvard.edu/abs/2014MNRAS.439.2542G} {439, 2542}

\bibitem[\protect\citeauthoryear{{Galliano}, {Madden}, {Jones}, {Wilson},
  {Bernard}  \& {Le Peintre}}{{Galliano} et~al.}{2003}]{2003A&A...407..159G}
{Galliano} F.,  {Madden} S.~C.,  {Jones} A.~P.,  {Wilson} C.~D.,  {Bernard}
  J.~P.,   {Le Peintre} F.,  2003, \mn@doi [\aap] {10.1051/0004-6361:20030814},
  \href {https://ui.adsabs.harvard.edu/abs/2003A&A...407..159G} {407, 159}

\bibitem[\protect\citeauthoryear{{Galliano}, {Madden}, {Jones}, {Wilson}  \&
  {Bernard}}{{Galliano} et~al.}{2005}]{2005A&A...434..867G}
{Galliano} F.,  {Madden} S.~C.,  {Jones} A.~P.,  {Wilson} C.~D.,   {Bernard}
  J.~P.,  2005, \mn@doi [\aap] {10.1051/0004-6361:20042369}, \href
  {https://ui.adsabs.harvard.edu/abs/2005A&A...434..867G} {434, 867}

\bibitem[\protect\citeauthoryear{{Garcia}}{{Garcia}}{1993}]{1993A&AS..100...47G}
{Garcia} A.~M.,  1993, \aaps, \href
  {http://adsabs.harvard.edu/abs/1993A%26AS..100...47G} {100, 47}

\bibitem[\protect\citeauthoryear{{Garn}, {Green}, {Riley}  \&
  {Alexander}}{{Garn} et~al.}{2009}]{2009MNRAS.397.1101G}
{Garn} T.,  {Green} D.~A.,  {Riley} J.~M.,   {Alexander} P.,  2009, \mn@doi
  [\mnras] {10.1111/j.1365-2966.2009.15073.x}, \href
  {http://adsabs.harvard.edu/abs/2009MNRAS.397.1101G} {397, 1101}

\bibitem[\protect\citeauthoryear{{Gil de Paz} et~al.,}{{Gil de Paz}
  et~al.}{2007}]{2007ApJS..173..185G}
{Gil de Paz} A.,  et~al., 2007, \mn@doi [\apjs] {10.1086/516636}, \href
  {https://ui.adsabs.harvard.edu/abs/2007ApJS..173..185G} {173, 185}

\bibitem[\protect\citeauthoryear{{Gordon} et~al.,}{{Gordon}
  et~al.}{2010}]{2010A&A...518L..89G}
{Gordon} K.~D.,  et~al., 2010, \mn@doi [\aap] {10.1051/0004-6361/201014541},
  \href {https://ui.adsabs.harvard.edu/abs/2010A&A...518L..89G} {518, L89}

\bibitem[\protect\citeauthoryear{{Gould} \& {Salpeter}}{{Gould} \&
  {Salpeter}}{1963}]{1963ApJ...138..393G}
{Gould} R.~J.,  {Salpeter} E.~E.,  1963, \mn@doi [\apj] {10.1086/147654}, \href
  {https://ui.adsabs.harvard.edu/abs/1963ApJ...138..393G} {138, 393}

\bibitem[\protect\citeauthoryear{{Hao}, {Kennicutt}, {Johnson}, {Calzetti},
  {Dale}  \& {Moustakas}}{{Hao} et~al.}{2011}]{2011ApJ...741..124H}
{Hao} C.-N.,  {Kennicutt} R.~C.,  {Johnson} B.~D.,  {Calzetti} D.,  {Dale}
  D.~A.,   {Moustakas} J.,  2011, \mn@doi [\apj] {10.1088/0004-637X/741/2/124},
  \href {https://ui.adsabs.harvard.edu/abs/2011ApJ...741..124H} {741, 124}

\bibitem[\protect\citeauthoryear{{Helou}, {Soifer}  \&
  {Rowan-Robinson}}{{Helou} et~al.}{1985}]{1985ApJ...298L...7H}
{Helou} G.,  {Soifer} B.~T.,   {Rowan-Robinson} M.,  1985, \mn@doi [\apjl]
  {10.1086/184556}, \href {http://adsabs.harvard.edu/abs/1985ApJ...298L...7H}
  {298, L7}

\bibitem[\protect\citeauthoryear{{Hippelein}, {Haas}, {Lemke}, {Stickel},
  {Tuffs}, {Klaas}  \& {V{\"o}lk}}{{Hippelein}
  et~al.}{2000}]{2000immm.proc...81H}
{Hippelein} A.,  {Haas} M.,  {Lemke} D.,  {Stickel} M.,  {Tuffs} R.,  {Klaas}
  U.,   {V{\"o}lk} H.,  2000, in {Berkhuijsen} E.~M.,  {Beck} R.,   {Walterbos}
  R.~A.~M.,  eds, Proceedings 232. WE-Heraeus Seminar. pp 81--84

\bibitem[\protect\citeauthoryear{{Hollenbach} \& {McKee}}{{Hollenbach} \&
  {McKee}}{1979}]{1979ApJS...41..555H}
{Hollenbach} D.,  {McKee} C.~F.,  1979, \mn@doi [\apjs] {10.1086/190631}, \href
  {https://ui.adsabs.harvard.edu/abs/1979ApJS...41..555H} {41, 555}

\bibitem[\protect\citeauthoryear{{Indebetouw} et~al.,}{{Indebetouw}
  et~al.}{2005}]{2005ApJ...619..931I}
{Indebetouw} R.,  et~al., 2005, \mn@doi [\apj] {10.1086/426679}, \href
  {http://adsabs.harvard.edu/abs/2005ApJ...619..931I} {619, 931}

\bibitem[\protect\citeauthoryear{{Inoue}}{{Inoue}}{2010}]{2010MNRAS.401.1325I}
{Inoue} A.~K.,  2010, \mn@doi [\mnras] {10.1111/j.1365-2966.2009.15730.x},
  \href {http://adsabs.harvard.edu/abs/2010MNRAS.401.1325I} {401, 1325}

\bibitem[\protect\citeauthoryear{{Inoue}}{{Inoue}}{2011}]{2011MNRAS.415.2920I}
{Inoue} A.~K.,  2011, \mn@doi [\mnras] {10.1111/j.1365-2966.2011.18906.x},
  \href {http://adsabs.harvard.edu/abs/2011MNRAS.415.2920I} {415, 2920}

\bibitem[\protect\citeauthoryear{{Israel}}{{Israel}}{1997}]{1997A&A...328..471I}
{Israel} F.~P.,  1997, \aap, \href
  {https://ui.adsabs.harvard.edu/abs/1997A&A...328..471I} {328, 471}

\bibitem[\protect\citeauthoryear{{Izotov}, {Guseva}, {Fricke}, {Kr{\"u}gel}  \&
  {Henkel}}{{Izotov} et~al.}{2014}]{2014A&A...570A..97I}
{Izotov} Y.~I.,  {Guseva} N.~G.,  {Fricke} K.~J.,  {Kr{\"u}gel} E.,   {Henkel}
  C.,  2014, \mn@doi [\aap] {10.1051/0004-6361/201423539}, \href
  {https://ui.adsabs.harvard.edu/abs/2014A&A...570A..97I} {570, A97}

\bibitem[\protect\citeauthoryear{{Jarrett}, {Chester}, {Cutri}, {Schneider}  \&
  {Huchra}}{{Jarrett} et~al.}{2003}]{2003AJ....125..525J}
{Jarrett} T.~H.,  {Chester} T.,  {Cutri} R.,  {Schneider} S.~E.,   {Huchra}
  J.~P.,  2003, \mn@doi [\aj] {10.1086/345794}, \href
  {https://ui.adsabs.harvard.edu/abs/2003AJ....125..525J} {125, 525}

\bibitem[\protect\citeauthoryear{{Jarrett}, {Cluver}, {Brown}, {Dale}, {Tsai}
  \& {Masci}}{{Jarrett} et~al.}{2019}]{2019ApJS..245...25J}
{Jarrett} T.~H.,  {Cluver} M.~E.,  {Brown} M.~J.~I.,  {Dale} D.~A.,  {Tsai}
  C.~W.,   {Masci} F.,  2019, \mn@doi [\apjs] {10.3847/1538-4365/ab521a}, \href
  {https://ui.adsabs.harvard.edu/abs/2019ApJS..245...25J} {245, 25}

\bibitem[\protect\citeauthoryear{{Juvela}, {Mattila}, {Lemke}, {Klaas},
  {Leinert}  \& {Kiss}}{{Juvela} et~al.}{2009}]{2009A&A...500..763J}
{Juvela} M.,  {Mattila} K.,  {Lemke} D.,  {Klaas} U.,  {Leinert} C.,   {Kiss}
  C.,  2009, \mn@doi [\aap] {10.1051/0004-6361/200811351}, \href
  {https://ui.adsabs.harvard.edu/abs/2009A&A...500..763J} {500, 763}

\bibitem[\protect\citeauthoryear{{Karachentsev} \& {Kaisina}}{{Karachentsev} \&
  {Kaisina}}{2013}]{2013AJ....146...46K}
{Karachentsev} I.~D.,  {Kaisina} E.~I.,  2013, \mn@doi [\aj]
  {10.1088/0004-6256/146/3/46}, \href
  {https://ui.adsabs.harvard.edu/abs/2013AJ....146...46K} {146, 46}

\bibitem[\protect\citeauthoryear{{Kelly}, {Shetty}, {Stutz}, {Kauffmann},
  {Goodman}  \& {Launhardt}}{{Kelly} et~al.}{2012}]{2012ApJ...752...55K}
{Kelly} B.~C.,  {Shetty} R.,  {Stutz} A.~M.,  {Kauffmann} J.,  {Goodman} A.~A.,
    {Launhardt} R.,  2012, \mn@doi [\apj] {10.1088/0004-637X/752/1/55}, \href
  {https://ui.adsabs.harvard.edu/abs/2012ApJ...752...55K} {752, 55}

\bibitem[\protect\citeauthoryear{{Kennicutt}}{{Kennicutt}}{1983}]{1983ApJ...272...54K}
{Kennicutt} R.~C. J.,  1983, \mn@doi [\apj] {10.1086/161261}, \href
  {https://ui.adsabs.harvard.edu/#abs/1983ApJ...272...54K} {272, 54}

\bibitem[\protect\citeauthoryear{{Kennicutt}}{{Kennicutt}}{1998a}]{1998ARA&A..36..189K}
{Kennicutt} Robert~C. J.,  1998a, \mn@doi [\araa]
  {10.1146/annurev.astro.36.1.189}, \href
  {https://ui.adsabs.harvard.edu/abs/1998ARA&A..36..189K} {36, 189}

\bibitem[\protect\citeauthoryear{{Kennicutt}}{{Kennicutt}}{1998b}]{1998ApJ...498..541K}
{Kennicutt} Robert~C. J.,  1998b, \mn@doi [\apj] {10.1086/305588}, \href
  {https://ui.adsabs.harvard.edu/abs/1998ApJ...498..541K} {498, 541}

\bibitem[\protect\citeauthoryear{{Kennicutt} \& {Evans}}{{Kennicutt} \&
  {Evans}}{2012}]{2012ARA&A..50..531K}
{Kennicutt} R.~C.,  {Evans} N.~J.,  2012, \mn@doi [\araa]
  {10.1146/annurev-astro-081811-125610}, \href
  {https://ui.adsabs.harvard.edu/abs/2012ARA&A..50..531K} {50, 531}

\bibitem[\protect\citeauthoryear{{Kewley}, {Geller}, {Jansen}  \&
  {Dopita}}{{Kewley} et~al.}{2002}]{2002AJ....124.3135K}
{Kewley} L.~J.,  {Geller} M.~J.,  {Jansen} R.~A.,   {Dopita} M.~A.,  2002,
  \mn@doi [\aj] {10.1086/344487}, \href
  {https://ui.adsabs.harvard.edu/abs/2002AJ....124.3135K} {124, 3135}

\bibitem[\protect\citeauthoryear{{Kirkpatrick} et~al.,}{{Kirkpatrick}
  et~al.}{2013}]{2013ApJ...778...51K}
{Kirkpatrick} A.,  et~al., 2013, \mn@doi [\apj] {10.1088/0004-637X/778/1/51},
  \href {http://adsabs.harvard} {778, 51}

\bibitem[\protect\citeauthoryear{{Koribalski}}{{Koribalski}}{1996}]{1996ASPC..106..238K}
{Koribalski} B.,  1996, in {Skillman} E.~D.,  ed.,  Astronomical Society of the
  Pacific Conference Series Vol. 106, The Minnesota Lectures on Extragalactic
  Neutral Hydrogen. p.~238

\bibitem[\protect\citeauthoryear{{Krugel}, {Siebenmorgen}, {Zota}  \&
  {Chini}}{{Krugel} et~al.}{1998}]{1998A&A...331L...9K}
{Krugel} E.,  {Siebenmorgen} R.,  {Zota} V.,   {Chini} R.,  1998, \aap, \href
  {https://ui.adsabs.harvard.edu/abs/1998A&A...331L...9K} {331, L9}

\bibitem[\protect\citeauthoryear{{Lacki} \& {Thompson}}{{Lacki} \&
  {Thompson}}{2010}]{2010ApJ...717..196L}
{Lacki} B.~C.,  {Thompson} T.~A.,  2010, \mn@doi [\apj]
  {10.1088/0004-637X/717/1/196}, \href
  {https://ui.adsabs.harvard.edu/abs/2010ApJ...717..196L} {717, 196}

\bibitem[\protect\citeauthoryear{{Lagache}, {Puget}  \& {Dole}}{{Lagache}
  et~al.}{2005}]{2005ARA&A..43..727L}
{Lagache} G.,  {Puget} J.-L.,   {Dole} H.,  2005, \mn@doi [\araa]
  {10.1146/annurev.astro.43.072103.150606}, \href
  {https://ui.adsabs.harvard.edu/abs/2005ARA&A..43..727L} {43, 727}

\bibitem[\protect\citeauthoryear{{Lal} \& {Rao}}{{Lal} \&
  {Rao}}{2007}]{2007MNRAS.374.1085L}
{Lal} D.~V.,  {Rao} A.~P.,  2007, \mn@doi [\mnras]
  {10.1111/j.1365-2966.2006.11225.x}, \href
  {https://ui.adsabs.harvard.edu/abs/2007MNRAS.374.1085L} {374, 1085}

\bibitem[\protect\citeauthoryear{{Lanz} et~al.,}{{Lanz}
  et~al.}{2013}]{2013ApJ...768...90L}
{Lanz} L.,  et~al., 2013, \mn@doi [\apj] {10.1088/0004-637X/768/1/90}, \href
  {http://adsabs.harvard.edu/abs/2013ApJ...768...90L} {768, 90}

\bibitem[\protect\citeauthoryear{{Lauberts} \& {Valentijn}}{{Lauberts} \&
  {Valentijn}}{1989}]{1989spce.book.....L}
{Lauberts} A.,  {Valentijn} E.~A.,  1989, {The surface photometry catalogue of
  the ESO-Uppsala galaxies}

\bibitem[\protect\citeauthoryear{{Laureijs}, {Klaas}, {Richards}, {Schulz}  \&
  {Abraham}}{{Laureijs} et~al.}{2003}]{2003sws..bookQ....L}
{Laureijs} R.~J.,  {Klaas} U.,  {Richards} P.~J.,  {Schulz} B.,   {Abraham} P.,
   2003, {The ISO Handbook, Volume IV - PHT - The Imaging Photo-Polarimeter}

\bibitem[\protect\citeauthoryear{{Leitherer}, {Calzetti}  \&
  {Martins}}{{Leitherer} et~al.}{2002}]{2002ApJ...574..114L}
{Leitherer} C.,  {Calzetti} D.,   {Martins} L.~P.,  2002, \mn@doi [\apj]
  {10.1086/340902}, \href {http://adsabs.harvard.edu/abs/2002ApJ...574..114L}
  {574, 114}

\bibitem[\protect\citeauthoryear{{Li} \& {Greenberg}}{{Li} \&
  {Greenberg}}{2003}]{2003ssac.proc...37L}
{Li} A.,  {Greenberg} J.~M.,  2003, in {Pirronello} V.,  {Krelowski} J.,
  {Manic{\`o}} G.,  eds,  Vol. 120, Solid State Astrochemistry. pp 37--84
  (\mn@eprint {arXiv} {astro-ph/0204392})

\bibitem[\protect\citeauthoryear{{Lianou}, {Barmby}, {Mosenkov}, {Lehnert}  \&
  {Karczewski}}{{Lianou} et~al.}{2019}]{2019arXiv190602712L}
{Lianou} S.,  {Barmby} P.,  {Mosenkov} A.,  {Lehnert} M.,   {Karczewski} O.,
  2019, arXiv e-prints, \href
  {https://ui.adsabs.harvard.edu/abs/2019arXiv190602712L} {p. arXiv:1906.02712}

\bibitem[\protect\citeauthoryear{{Lira}, {Videla}, {Wu}, {Alonso-Herrero},
  {Alexander}  \& {Ward}}{{Lira} et~al.}{2013}]{2013ApJ...764..159L}
{Lira} P.,  {Videla} L.,  {Wu} Y.,  {Alonso-Herrero} A.,  {Alexander} D.~M.,
  {Ward} M.,  2013, \mn@doi [\apj] {10.1088/0004-637X/764/2/159}, \href
  {https://ui.adsabs.harvard.edu/abs/2013ApJ...764..159L} {764, 159}

\bibitem[\protect\citeauthoryear{{Magnelli} et~al.,}{{Magnelli}
  et~al.}{2015}]{2015A&A...573A..45M}
{Magnelli} B.,  et~al., 2015, \mn@doi [\aap] {10.1051/0004-6361/201424937},
  \href {https://ui.adsabs.harvard.edu/abs/2015A&A...573A..45M} {573, A45}

\bibitem[\protect\citeauthoryear{{Marleau}, {Noriega-Crespo}, {Rieke}  \& {MIPS
  Team}}{{Marleau} et~al.}{2004}]{2004AAS...20514108M}
{Marleau} F.~R.,  {Noriega-Crespo} A.,  {Rieke} G.,   {MIPS Team} 2004, in
  American Astronomical Society Meeting Abstracts. p.~1580

\bibitem[\protect\citeauthoryear{{Marleau} et~al.,}{{Marleau}
  et~al.}{2006}]{2006ApJ...646..929M}
{Marleau} F.~R.,  et~al., 2006, \mn@doi [\apj] {10.1086/504975}, \href
  {https://ui.adsabs.harvard.edu/abs/2006ApJ...646..929M} {646, 929}

\bibitem[\protect\citeauthoryear{{Martin} et~al.,}{{Martin}
  et~al.}{2005}]{2005ApJ...619L...1M}
{Martin} D.~C.,  et~al., 2005, \mn@doi [\apj] {10.1086/426387}, \href
  {https://ui.adsabs.harvard.edu/abs/2005ApJ...619L...1M} {619, L1}

\bibitem[\protect\citeauthoryear{{Mathewson} \& {Ford}}{{Mathewson} \&
  {Ford}}{1996}]{1996ApJS..107...97M}
{Mathewson} D.~S.,  {Ford} V.~L.,  1996, \mn@doi [\apjs] {10.1086/192356},
  \href {https://ui.adsabs.harvard.edu/abs/1996ApJS..107...97M} {107, 97}

\bibitem[\protect\citeauthoryear{{Matthews}, {McCutcheon}, {Kirk}, {White}  \&
  {Cohen}}{{Matthews} et~al.}{2008}]{2008AJ....136.2083M}
{Matthews} H.~E.,  {McCutcheon} W.~H.,  {Kirk} H.,  {White} G.~J.,   {Cohen}
  M.,  2008, \mn@doi [\aj] {10.1088/0004-6256/136/5/2083}, \href
  {http://adsabs.harvard.edu/abs/2008AJ....136.2083M} {136, 2083}

\bibitem[\protect\citeauthoryear{{Mauch}, {Murphy}, {Buttery}, {Curran},
  {Hunstead}, {Piestrzynski}, {Robertson}  \& {Sadler}}{{Mauch}
  et~al.}{2003}]{2003MNRAS.342.1117M}
{Mauch} T.,  {Murphy} T.,  {Buttery} H.~J.,  {Curran} J.,  {Hunstead} R.~W.,
  {Piestrzynski} B.,  {Robertson} J.~G.,   {Sadler} E.~M.,  2003, \mn@doi
  [\mnras] {10.1046/j.1365-8711.2003.06605.x}, \href
  {http://adsabs.harvard.edu/abs/2003MNRAS.342.1117M} {342, 1117}

\bibitem[\protect\citeauthoryear{{Misselt}, {Gordon}, {Clayton}  \&
  {Wolff}}{{Misselt} et~al.}{2001}]{2001ApJ...551..277M}
{Misselt} K.~A.,  {Gordon} K.~D.,  {Clayton} G.~C.,   {Wolff} M.~J.,  2001,
  \mn@doi [\apj] {10.1086/320083}, \href
  {http://adsabs.harvard.edu/abs/2001ApJ...551..277M} {551, 277}

\bibitem[\protect\citeauthoryear{{Mocanu} et~al.,}{{Mocanu}
  et~al.}{2013}]{2013ApJ...779...61M}
{Mocanu} L.~M.,  et~al., 2013, \mn@doi [\apj] {10.1088/0004-637X/779/1/61},
  \href {https://ui.adsabs.harvard.edu/abs/2013ApJ...779...61M} {779, 61}

\bibitem[\protect\citeauthoryear{{Mu{\~n}oz-Mateos}, {Gil de Paz}, {Boissier},
  {Zamorano}, {Jarrett}, {Gallego}  \& {Madore}}{{Mu{\~n}oz-Mateos}
  et~al.}{2007}]{2007ApJ...658.1006M}
{Mu{\~n}oz-Mateos} J.~C.,  {Gil de Paz} A.,  {Boissier} S.,  {Zamorano} J.,
  {Jarrett} T.,  {Gallego} J.,   {Madore} B.~F.,  2007, \mn@doi [\apj]
  {10.1086/511812}, \href {http://adsabs.harvard.edu/abs/2007ApJ...658.1006M}
  {658, 1006}

\bibitem[\protect\citeauthoryear{{Murakami} et~al.,}{{Murakami}
  et~al.}{2007}]{2007PASJ...59S.369M}
{Murakami} H.,  et~al., 2007, \mn@doi [\pasj] {10.1093/pasj/59.sp2.S369}, \href
  {http://adsabs.harvard.edu/abs/2007PASJ...59S.369M} {59, S369}

\bibitem[\protect\citeauthoryear{{Murphy} et~al.,}{{Murphy}
  et~al.}{2010}]{2010ApJ...709L.108M}
{Murphy} E.~J.,  et~al., 2010, \mn@doi [\apjl] {10.1088/2041-8205/709/2/L108},
  \href {https://ui.adsabs.harvard.edu/abs/2010ApJ...709L.108M} {709, L108}

\bibitem[\protect\citeauthoryear{{Netzer}}{{Netzer}}{2015}]{2015ARA&A..53..365N}
{Netzer} H.,  2015, \mn@doi [\araa] {10.1146/annurev-astro-082214-122302},
  \href {https://ui.adsabs.harvard.edu/abs/2015ARA&A..53..365N} {53, 365}

\bibitem[\protect\citeauthoryear{{Neugebauer} et~al.,}{{Neugebauer}
  et~al.}{1984}]{1984ApJ...278L...1N}
{Neugebauer} G.,  et~al., 1984, \mn@doi [\apjl] {10.1086/184209}, \href
  {https://ui.adsabs.harvard.edu/abs/1984ApJ...278L...1N} {278, L1}

\bibitem[\protect\citeauthoryear{{Pagani}, {Lef{\`e}vre}, {Juvela}, {Pelkonen}
  \& {Schuller}}{{Pagani} et~al.}{2015}]{2015A&A...574L...5P}
{Pagani} L.,  {Lef{\`e}vre} C.,  {Juvela} M.,  {Pelkonen} V.-M.,   {Schuller}
  F.,  2015, \mn@doi [\aap] {10.1051/0004-6361/201425095}, \href
  {http://adsabs.harvard.edu/abs/2015A%26A...574L...5P} {574, L5}

\bibitem[\protect\citeauthoryear{{Pilbratt} et~al.,}{{Pilbratt}
  et~al.}{2010}]{2010A&A...518L...1P}
{Pilbratt} G.~L.,  et~al., 2010, \mn@doi [\aap] {10.1051/0004-6361/201014759},
  \href {http://adsabs.harvard.edu/abs/2010A%26A...518L...1P} {518, L1}

\bibitem[\protect\citeauthoryear{{Planck Collaboration} \& {Lawrence}}{{Planck
  Collaboration} \& {Lawrence}}{2011}]{2011AAS...21724301P}
{Planck Collaboration} {Lawrence} C.~R.,  2011, in American Astronomical
  Society Meeting Abstracts \#217. p. 243.01

\bibitem[\protect\citeauthoryear{{Planck Collaboration} et~al.,}{{Planck
  Collaboration} et~al.}{2014}]{refId0}
{Planck Collaboration} et~al., 2014, \mn@doi [A\&A]
  {10.1051/0004-6361/201321529}, 571, A1

\bibitem[\protect\citeauthoryear{{Popescu}, {Tuffs}, {Dopita}, {Fischera},
  {Kylafis}  \& {Madore}}{{Popescu} et~al.}{2011}]{2011A&A...527A.109P}
{Popescu} C.~C.,  {Tuffs} R.~J.,  {Dopita} M.~A.,  {Fischera} J.,  {Kylafis}
  N.~D.,   {Madore} B.~F.,  2011, \mn@doi [\aap] {10.1051/0004-6361/201015217},
  \href {https://ui.adsabs.harvard.edu/abs/2011A&A...527A.109P} {527, A109}

\bibitem[\protect\citeauthoryear{{Reach} et~al.,}{{Reach}
  et~al.}{1995}]{1995ApJ...451..188R}
{Reach} W.~T.,  et~al., 1995, \mn@doi [\apj] {10.1086/176210}, \href
  {https://ui.adsabs.harvard.edu/abs/1995ApJ...451..188R} {451, 188}

\bibitem[\protect\citeauthoryear{{Read} et~al.,}{{Read}
  et~al.}{2018}]{2018MNRAS.480.5625R}
{Read} S.~C.,  et~al., 2018, \mn@doi [\mnras] {10.1093/mnras/sty2198}, \href
  {https://ui.adsabs.harvard.edu/abs/2018MNRAS.480.5625R} {480, 5625}

\bibitem[\protect\citeauthoryear{{Rickard} \& {Harvey}}{{Rickard} \&
  {Harvey}}{1984}]{1984AJ.....89.1520R}
{Rickard} L.~J.,  {Harvey} P.~M.,  1984, \mn@doi [\aj] {10.1086/113652}, \href
  {https://ui.adsabs.harvard.edu/abs/1984AJ.....89.1520R} {89, 1520}

\bibitem[\protect\citeauthoryear{{Rowan-Robinson} et~al.,}{{Rowan-Robinson}
  et~al.}{2008}]{2008ASPC..381..216R}
{Rowan-Robinson} M.,  et~al., 2008, in {Chary} R.-R.,  {Teplitz} H.~I.,
  {Sheth} K.,  eds,  Astronomical Society of the Pacific Conference Series Vol.
  381, Infrared Diagnostics of Galaxy Evolution. p.~216 (\mn@eprint {}
  {astro-ph/0603737})

\bibitem[\protect\citeauthoryear{{Rowan-Robinson} et~al.,}{{Rowan-Robinson}
  et~al.}{2010}]{2010MNRAS.409....2R}
{Rowan-Robinson} M.,  et~al., 2010, \mn@doi [\mnras]
  {10.1111/j.1365-2966.2010.17041.x}, \href
  {http://adsabs.harvard.edu/abs/2010MNRAS.409....2R} {409, 2}

\bibitem[\protect\citeauthoryear{{Rowlands}, {Gomez}, {Dunne},
  {Arag{\'o}n-Salamanca}, {Dye}, {Maddox}, {da Cunha}  \& {van der
  Werf}}{{Rowlands} et~al.}{2014}]{2014MNRAS.441.1040R}
{Rowlands} K.,  {Gomez} H.~L.,  {Dunne} L.,  {Arag{\'o}n-Salamanca} A.,  {Dye}
  S.,  {Maddox} S.,  {da Cunha} E.,   {van der Werf} P.,  2014, \mn@doi
  [\mnras] {10.1093/mnras/stu605}, \href
  {http://adsabs.harvard.edu/abs/2014MNRAS.441.1040R} {441, 1040}

\bibitem[\protect\citeauthoryear{{Sadler} \& {Hunstead}}{{Sadler} \&
  {Hunstead}}{2001}]{2001ASPC..232...53S}
{Sadler} E.~M.,  {Hunstead} R.~W.,  2001, {SUMSS: Wide-field Radio Imaging of
  the Southern Sky.}.
p.~53

\bibitem[\protect\citeauthoryear{{Salpeter}}{{Salpeter}}{1955}]{1955ApJ...121..161S}
{Salpeter} E.~E.,  1955, \mn@doi [\apj] {10.1086/145971}, \href
  {http://adsabs.harvard.edu/abs/1955ApJ...121..161S} {121, 161}

\bibitem[\protect\citeauthoryear{{S{\'a}nchez-Monge}, {Kurtz}, {Palau},
  {Estalella}, {Shepherd}, {Lizano}, {Franco}  \& {Garay}}{{S{\'a}nchez-Monge}
  et~al.}{2013}]{2013ApJ...766..114S}
{S{\'a}nchez-Monge} {\'A}.,  {Kurtz} S.,  {Palau} A.,  {Estalella} R.,
  {Shepherd} D.,  {Lizano} S.,  {Franco} J.,   {Garay} G.,  2013, \mn@doi
  [\apj] {10.1088/0004-637X/766/2/114}, \href
  {https://ui.adsabs.harvard.edu/abs/2013ApJ...766..114S} {766, 114}

\bibitem[\protect\citeauthoryear{{Sanders} \& {Mirabel}}{{Sanders} \&
  {Mirabel}}{1996}]{1996ARA&A..34..749S}
{Sanders} D.~B.,  {Mirabel} I.~F.,  1996, \mn@doi [\araa]
  {10.1146/annurev.astro.34.1.749}, \href
  {https://ui.adsabs.harvard.edu/abs/1996ARA&A..34..749S} {34, 749}

\bibitem[\protect\citeauthoryear{{Schmitt}, {Calzetti}, {Armus}, {Giavalisco},
  {Heckman}, {Kennicutt}, {Leitherer}  \& {Meurer}}{{Schmitt}
  et~al.}{2006}]{2006ApJ...643..173S}
{Schmitt} H.~R.,  {Calzetti} D.,  {Armus} L.,  {Giavalisco} M.,  {Heckman}
  T.~M.,  {Kennicutt} R.~C. J.,  {Leitherer} C.,   {Meurer} G.~R.,  2006,
  \mn@doi [\apj] {10.1086/501512}, \href
  {https://ui.adsabs.harvard.edu/#abs/2006ApJ...643..173S} {643, 173}

\bibitem[\protect\citeauthoryear{{Scoville} et~al.,}{{Scoville}
  et~al.}{2014}]{2014ApJ...783...84S}
{Scoville} N.,  et~al., 2014, \mn@doi [\apj] {10.1088/0004-637X/783/2/84},
  \href {https://ui.adsabs.harvard.edu/abs/2014ApJ...783...84S} {783, 84}

\bibitem[\protect\citeauthoryear{{Seymour}, {Huynh}, {Dwelly}, {Symeonidis},
  {Hopkins}, {McHardy}, {Page}  \& {Rieke}}{{Seymour}
  et~al.}{2009}]{2009MNRAS.398.1573S}
{Seymour} N.,  {Huynh} M.,  {Dwelly} T.,  {Symeonidis} M.,  {Hopkins} A.,
  {McHardy} I.~M.,  {Page} M.~J.,   {Rieke} G.,  2009, \mn@doi [\mnras]
  {10.1111/j.1365-2966.2009.15224.x}, \href
  {https://ui.adsabs.harvard.edu/abs/2009MNRAS.398.1573S} {398, 1573}

\bibitem[\protect\citeauthoryear{{Singh} et~al.,}{{Singh}
  et~al.}{2014}]{2014SPIE.9144E..1SS}
{Singh} K.~P.,  et~al., 2014, in \procspie. p. 91441S,
  \mn@doi{10.1117/12.2062667}

\bibitem[\protect\citeauthoryear{{Skrutskie} et~al.,}{{Skrutskie}
  et~al.}{2006}]{2006AJ....131.1163S}
{Skrutskie} M.~F.,  et~al., 2006, \mn@doi [\aj] {10.1086/498708}, \href
  {https://ui.adsabs.harvard.edu/abs/2006AJ....131.1163S} {131, 1163}

\bibitem[\protect\citeauthoryear{{Spinoglio}, {Andreani}  \&
  {Malkan}}{{Spinoglio} et~al.}{2002}]{2002ApJ...572..105S}
{Spinoglio} L.,  {Andreani} P.,   {Malkan} M.~A.,  2002, \mn@doi [\apj]
  {10.1086/340302}, \href
  {https://ui.adsabs.harvard.edu/abs/2002ApJ...572..105S} {572, 105}

\bibitem[\protect\citeauthoryear{{Stark}, {Davidson}, {Platt}, {Harper},
  {Pernic}, {Loewenstein}, {Engargiola}  \& {Casey}}{{Stark}
  et~al.}{1989}]{1989ApJ...337..650S}
{Stark} A.~A.,  {Davidson} J.~A.,  {Platt} S.,  {Harper} D.~A.,  {Pernic} R.,
  {Loewenstein} R.,  {Engargiola} G.,   {Casey} S.,  1989, \mn@doi [\apj]
  {10.1086/167136}, \href
  {https://ui.adsabs.harvard.edu/abs/1989ApJ...337..650S} {337, 650}

\bibitem[\protect\citeauthoryear{{Story} et~al.,}{{Story}
  et~al.}{2013}]{2013ApJ...779...86S}
{Story} K.~T.,  et~al., 2013, \mn@doi [\apj] {10.1088/0004-637X/779/1/86},
  \href {https://ui.adsabs.harvard.edu/abs/2013ApJ...779...86S} {779, 86}

\bibitem[\protect\citeauthoryear{{Subramaniam} et~al.,}{{Subramaniam}
  et~al.}{2016}]{2016ApJ...833L..27S}
{Subramaniam} A.,  et~al., 2016, \mn@doi [\apjl] {10.3847/2041-8213/833/2/L27},
  \href {https://ui.adsabs.harvard.edu/abs/2016ApJ...833L..27S} {833, L27}

\bibitem[\protect\citeauthoryear{{Swarup}, {Ananthakrishnan}, {Kapahi}, {Rao},
  {Subrahmanya}  \& {Kulkarni}}{{Swarup} et~al.}{1991}]{1991CuSc...60...95S}
{Swarup} G.,  {Ananthakrishnan} S.,  {Kapahi} V.~K.,  {Rao} A.~P.,
  {Subrahmanya} C.~R.,   {Kulkarni} V.~K.,  1991, Current Science, Vol.~60,
  NO.2/JAN25, P.~95, 1991, \href
  {http://adsabs.harvard.edu/abs/1991CuSc...60...95S} {60, 95}

\bibitem[\protect\citeauthoryear{{Tabatabaei}, {Wei{\ss}}, {Combes}, {Henkel},
  {Menten}, {Beck}, {Kov{\'a}cs}  \& {G{\"u}sten}}{{Tabatabaei}
  et~al.}{2013}]{2013A&A...555A.128T}
{Tabatabaei} F.~S.,  {Wei{\ss}} A.,  {Combes} F.,  {Henkel} C.,  {Menten}
  K.~M.,  {Beck} R.,  {Kov{\'a}cs} A.,   {G{\"u}sten} R.,  2013, \mn@doi [\aap]
  {10.1051/0004-6361/201321487}, \href
  {http://adsabs.harvard.edu/abs/2013A%26A...555A.128T} {555, A128}

\bibitem[\protect\citeauthoryear{{Tandon} et~al.,}{{Tandon}
  et~al.}{2017a}]{2017JApA...38...28T}
{Tandon} S.~N.,  et~al., 2017a, \mn@doi [Journal of Astrophysics and Astronomy]
  {10.1007/s12036-017-9445-x}, \href
  {http://adsabs.harvard.edu/abs/2017JApA...38...28T} {38, 28}

\bibitem[\protect\citeauthoryear{{Tandon} et~al.,}{{Tandon}
  et~al.}{2017b}]{2017AJ....154..128T}
{Tandon} S.~N.,  et~al., 2017b, \mn@doi [\aj] {10.3847/1538-3881/aa8451}, \href
  {http://adsabs.harvard.edu/abs/2017AJ....154..128T} {154, 128}

\bibitem[\protect\citeauthoryear{{Thilker} et~al.,}{{Thilker}
  et~al.}{2007}]{2007ApJS..173..538T}
{Thilker} D.~A.,  et~al., 2007, \mn@doi [\apjs] {10.1086/523853}, \href
  {http://adsabs.harvard.edu/abs/2007ApJS..173..538T} {173, 538}

\bibitem[\protect\citeauthoryear{{Thuma}, {Neininger}, {Klein}  \&
  {Wielebinski}}{{Thuma} et~al.}{2000}]{2000A&A...358...65T}
{Thuma} G.,  {Neininger} N.,  {Klein} U.,   {Wielebinski} R.,  2000, \aap,
  \href {http://adsabs.harvard.edu/abs/2000A%26A...358...65T} {358, 65}

\bibitem[\protect\citeauthoryear{{Vieira} et~al.,}{{Vieira}
  et~al.}{2010}]{2010ApJ...719..763V}
{Vieira} J.~D.,  et~al., 2010, \mn@doi [\apj] {10.1088/0004-637X/719/1/763},
  \href {https://ui.adsabs.harvard.edu/abs/2010ApJ...719..763V} {719, 763}

\bibitem[\protect\citeauthoryear{{Vlahakis}, {Eales}  \& {Dunne}}{{Vlahakis}
  et~al.}{2007}]{2007MNRAS.379.1042V}
{Vlahakis} C.,  {Eales} S.,   {Dunne} L.,  2007, \mn@doi [\mnras]
  {10.1111/j.1365-2966.2007.12007.x}, \href
  {https://ui.adsabs.harvard.edu/abs/2007MNRAS.379.1042V} {379, 1042}

\bibitem[\protect\citeauthoryear{{Werner} et~al.,}{{Werner}
  et~al.}{2004}]{2004AAS...204.3301W}
{Werner} M.,  et~al., 2004, in American Astronomical Society Meeting Abstracts
  \#204. p.~699

\bibitem[\protect\citeauthoryear{{Whitford}}{{Whitford}}{1958}]{1958AJ.....63..201W}
{Whitford} A.~E.,  1958, \mn@doi [\aj] {10.1086/107725}, \href
  {http://adsabs.harvard.edu/abs/1958AJ.....63..201W} {63, 201}

\bibitem[\protect\citeauthoryear{{Wild} et~al.,}{{Wild}
  et~al.}{2014}]{2014MNRAS.440.1880W}
{Wild} V.,  et~al., 2014, \mn@doi [\mnras] {10.1093/mnras/stu212}, \href
  {https://ui.adsabs.harvard.edu/abs/2014MNRAS.440.1880W} {440, 1880}

\bibitem[\protect\citeauthoryear{{Witt} \& {Gordon}}{{Witt} \&
  {Gordon}}{2000}]{2000ApJ...528..799W}
{Witt} A.~N.,  {Gordon} K.~D.,  2000, \mn@doi [\apj] {10.1086/308197}, \href
  {http://adsabs.harvard.edu/abs/2000ApJ...528..799W} {528, 799}

\bibitem[\protect\citeauthoryear{{Wolfire}, {Hollenbach}, {McKee}, {Tielens}
  \& {Bakes}}{{Wolfire} et~al.}{1995}]{1995ApJ...443..152W}
{Wolfire} M.~G.,  {Hollenbach} D.,  {McKee} C.~F.,  {Tielens} A.~G.~G.~M.,
  {Bakes} E.~L.~O.,  1995, \mn@doi [\apj] {10.1086/175510}, \href
  {https://ui.adsabs.harvard.edu/abs/1995ApJ...443..152W} {443, 152}

\bibitem[\protect\citeauthoryear{{Wright} et~al.,}{{Wright}
  et~al.}{2010}]{2010AJ....140.1868W}
{Wright} E.~L.,  et~al., 2010, \mn@doi [\aj] {10.1088/0004-6256/140/6/1868},
  \href {https://ui.adsabs.harvard.edu/abs/2010AJ....140.1868W} {140, 1868}

\bibitem[\protect\citeauthoryear{{Yamamura} et~al.,}{{Yamamura}
  et~al.}{2009}]{2009ASPC..418....3Y}
{Yamamura} I.,  et~al., 2009, in {Onaka} T.,  {White} G.~J.,  {Nakagawa} T.,
  {Yamamura} I.,  eds,  Astronomical Society of the Pacific Conference Series
  Vol. 418, AKARI, a Light to Illuminate the Misty Universe. p.~3

\bibitem[\protect\citeauthoryear{{Yun}, {Reddy}  \& {Condon}}{{Yun}
  et~al.}{2001}]{2001ApJ...554..803Y}
{Yun} M.~S.,  {Reddy} N.~A.,   {Condon} J.~J.,  2001, \mn@doi [\apj]
  {10.1086/323145}, \href
  {https://ui.adsabs.harvard.edu/abs/2001ApJ...554..803Y} {554, 803}

\bibitem[\protect\citeauthoryear{{van Dishoeck}}{{van
  Dishoeck}}{2004}]{2004ARA&A..42..119V}
{van Dishoeck} E.~F.,  2004, \mn@doi [\araa]
  {10.1146/annurev.astro.42.053102.134010}, \href
  {https://ui.adsabs.harvard.edu/abs/2004ARA&A..42..119V} {42, 119}

\bibitem[\protect\citeauthoryear{{van der Kruit}}{{van der
  Kruit}}{1971}]{1971A&A....15..110V}
{van der Kruit} P.~C.,  1971, \aap, \href
  {https://ui.adsabs.harvard.edu/abs/1971A&A....15..110V} {15, 110}

\makeatother
\end{thebibliography}



\appendix
\section{Multiwavelength Data for SED modeling}
\begin{table*}
\caption{Flux densities$^*$ and uncertainties (in mJy) in each band. }
\label{table:fluxes}
\begin{center}
\begin{tabular}{|c|r|r|r|r|r|}
\hline
\textbf{Bands} 				& \textbf{IC\,5325} 		& \textbf{NGC\,7496}     	& \textbf{NGC\,7590}          	& \textbf{NGC\,7599}        	& \textbf{Reference} 			\\\hline
\textit{\textbf{GALEX}}     &     						&      						&      							&       						& \cite{2007ApJS..173..185G}    \\
FUV (1516\,\AA)          	& 8.86 $\pm$ 0.07   		& 4.00 $\pm$ 0.02   		& -                             & -                             &                    			\\
NUV (2269\,\AA)      		& 13.29 $\pm$ 0.11  		& 5.97 $\pm$ 0.02   		& -                             & -                             &                    			\\\hline
\textbf{UVIT }              &     						&      						&      							&       						& This work	(Section\,\ref{uvit})		             \\
F154W (1541\,\AA)      		& -                 		& -    						& 8.49  $\pm$ 0.42              & 8.92  $\pm$ 0.45    			&                    			\\
F172M (1717\,\AA)    		& -                 		& -  						&   9.87 $\pm$ 0.49             & 10.78  $\pm$ 0.54             &                    			\\
N245M (2447\,\AA)     		& -                 		& -    						& 10.99  $\pm$ 0.55             & 12.70  $\pm$ 0.64     		&                    			\\
N279N (2792\,\AA)      		& -                 		& -  						&   13.44 $\pm$ 0.67            & 17.53  $\pm$ 0.88   			&                    			\\\hline
\textbf{ Bessel 90 }        &     						&      						&      							&       						&                     			\\
B (4363\,\AA)    			& 79.81 $\pm$ 6.44  		& 75.25 $\pm$ 6.29     		& 64.28 $\pm$ 5.24     			& 80.90 $\pm$ 6.59          	&  \cite{1989spce.book.....L}    \\
RC (6407\,\AA)      		& 163.90 $\pm$ 13.60		& 131.75 $\pm$ 11.10      	& 120.38 $\pm$ 9.99      		& 114.12 $\pm$ 9.52           	&  \cite{1989spce.book.....L}    \\
IC (7982\,\AA)      		& -                    		& -                      	& 169.41 $\pm$ 8.23      		& 170.50 $\pm$ 8.31             &  \cite{1996ApJS..107...97M}    \\\hline
\textbf{2MASS}              &     						&      						&      							&       						&  \cite{2003AJ....125..525J}    \\
J (1.25\,$\mu$m)     		& 332.00 $\pm$ 7.02     	& 257.00 $\pm$ 5.23      	& 305.00 $\pm$ 2.79      		& 291.00 $\pm$ 6.15            	&                    			\\
H (1.65\,$\mu$m)     		& 375.00 $\pm$ 8.33     	& 299.00 $\pm$ 6.94      	& 345.00 $\pm$ 3.79      		& 335.00 $\pm$ 8.05            	&                    			\\
K$_s$ (2.17\,$\mu$m)    	& 304.00 $\pm$ 9.29     	& 232.00 $\pm$ 9.34      	& 290.00 $\pm$ 4.56      		& 295.00 $\pm$ 9.03           	&                   			\\\hline
\textit{\textbf{WISE}}      &     						&      						&      							&       						&  \cite{2019ApJS..245...25J}   \\
3.4\,$\mu$m        			& 163.27 $\pm$ 9.90     	& 150.15 $\pm$ 9.10     	& 170.75 $\pm$ 10.40     		& 162.30 $\pm$ 11.60       		&                    			\\
4.6\,$\mu$m        			& 94.47 $\pm$ 5.70    		& 95.52 $\pm$ 5.80      	& 106.68 $\pm$ 6.50     		& 103.86 $\pm$ 6.30           	&                    			\\
12\,$\mu$m         			& 378.30 $\pm$ 38.00    	& 445.20 $\pm$ 44.80     	& 481.80 $\pm$ 48.50     		& 445.90 $\pm$ 44.80      		&                    			\\
22\,$\mu$m          		& 547.90 $\pm$ 55.20    	& 1617.70 $\pm$ 162.70   	& 743.40 $\pm$ 74.80     		& 648.10 $\pm$ 65.30      		&                    			\\\hline
\textit{\textbf{IRAC}}     	&     						&      						&      							&       						&  \cite{2004ApJS..154...10F}   \\
3.6\,$\mu$m         		& 161.56 $\pm$ 16.16   		& 146.27 $\pm$ 14.63     	& 177.60 $\pm$ 17.76     		& 155.50 $\pm$ 15.55        	&                    			\\
4.5\,$\mu$m         		& 102.23 $\pm$ 10.22   		& 100.14 $\pm$ 10.01      	& 114.28 $\pm$ 11.43     		& 103.29 $\pm$ 10.33      		&                   			\\
5.8\,$\mu$m          		& -   						& 250.28 $\pm$ 25.03     	& 298.74 $\pm$ 29.87     		& -          					&                    			\\
8.0\,$\mu$m          		& -   						& 592.80 $\pm$ 59.28   		& 826.70 $\pm$ 82.67     		& -        						&                    			\\\hline
\textit{\textbf{IRAS}}      &     						&      						&      							&       						&        NED            		\\
12\,$\mu$m            		& 390.0 $\pm$ 89.0    		& 580.0 $\pm$ 104.3       	& 690.0 $\pm$ 61.5        		& 740.0 $\pm$ 78.8            	&                    			\\
25\,$\mu$m         			& 470.0 $\pm$ 42.3    		& 1930.0 $\pm$ 139.3      	& 890.0 $\pm$ 75.8        		& 770.0 $\pm$ 81.3            	&                    			\\
60\,$\mu$m         			& 4530.0 $\pm$ 317.1  		& 10100.0 $\pm$ 909.8     	& 7690.0 $\pm$ 770.1       		& 6390.0 $\pm$ 704.5          	&                    			\\
100\,$\mu$m     			& 14500.0 $\pm$ 1015.0		& 16600.0 $\pm$ 832.7     	& 20800.0 $\pm$ 1459.3      	& 18300.0 $\pm$ 1284.0        	&                    			\\\hline
\textit{\textbf{AKARI}}     &     						&     						&      							&       						&  \cite{2009ASPC..418....3Y}   \\
N60 (65\,$\mu$m)           	& 10819.90 $\pm$ 2163.98	& 14325.40 $\pm$ 2865.08 	& 12243.00 $\pm$ 2448.60 		& 8429.10 $\pm$ 1685.82  		&                    			\\
Wide-S (90\,$\mu$m)         & 12011.90 $\pm$ 2402.38	& 16214.80 $\pm$ 3242.96  	& 19567.10 $\pm$ 3913.42  		& 17661.70 $\pm$ 3232.30    	&                    			\\
N160 (160\,$\mu$m)       	& 15642.90 $\pm$ 3128.58	& 8220.80 $\pm$ 1644.20  	& 19540.60 $\pm$ 3908.12 		& 17196.90  $\pm$ 3439.38    	&                    			\\\hline
\textit{\textbf{ISOPHOT$^{**}$}} &     					&      						&      							&       						& \cite{2002ApJ...572..105S}    \\
120\,$\mu$m    				& -     					& 20800 $\pm$ 4160   		& 25900 $\pm$ 5180     			& -                             &                         		\\            
150\,$\mu$m     			& -   						& 16200 $\pm$ 3240    		& 22800 $\pm$ 4560        		& -                             &                   			\\
170\,$\mu$m     			& -    						& 13900 $\pm$ 2780    		& 17600 $\pm$ 3520        		& -                             &                   			\\
180\,$\mu$m     			& -    						& 8890 $\pm$ 1778     		& 12400 $\pm$ 2480        		& -                             &                   			\\
200\,$\mu$m    				& -     					& 6250 $\pm$ 1250     		& 7910 $\pm$ 1582         		& -                             &                  				\\\hline
\textit{\textbf{Planck}} \textbf{HFI} &     			&      						&      							&   \multicolumn{2}{|r|}{\cite{refId0}}                			\\
857\,GHz (350\,$\mu$m)     	& 4610.8 $\pm$ 537.7        & 3917.8 $\pm$ 362.6  		& -                             & 6031.9 $\pm$ 761.3    		&                    			\\
545\,GHz (550\,$\mu$m)     	& 1459.6 $\pm$ 222.7        & 1139.0 $\pm$ 167.8  		& -                             & 2115.1 $\pm$ 200.2    		&                    			\\
353\,GHz (850\,$\mu$m)     	& 513.9 $\pm$ 89.8          & 323.8 $\pm$ 80.6    		& 651.3 $\pm$ 67.4    			& 651.3 $\pm$ 67.4       		&                    			\\\hline
\textbf{SPT}                &     						&      						&      							&      							& This work (Section\,\ref{spt})    \\
220\,GHz (1.4\,mm)       	& 108.02 $\pm$ 4.84      	& 64.61 $\pm$ 4.62  		& 100.95 $\pm$ 4.85             & 134.049 $\pm$ 5.40    		&                    			\\
150\,GHz (2.0\,mm)      	& 21.58 $\pm$ 1.35         	& 13.98 $\pm$ 1.25  		& 26.22 $\pm$ 1.30              & 31.26 $\pm$ 1.45   			&                    			\\
95\,GHz (3.2\,mm)       	& 14.16 $\pm$ 2.33        	& 6.59 $\pm$ 2.29    		& 13.10  $\pm$ 2.33   			& 15.33 $\pm$ 2.57     			&                    			\\\hline
\textbf{GMRT }              &     						&      						&      							&       						& This work (Section\,\ref{gmrt})               		\\
1300\,MHz (23\,cm)    		& -        					& 21.40 $\pm$ 3.10          & -                             & -                             &                    			\\
610\,MHz (49\,cm)      		& 66.00 $\pm$ 5.08         	& -                         & 112.00 $\pm$ 6.14             & 84.50 $\pm$ 5.34              &                    			\\
325\,MHz (92\,cm)      		& 91.00 $\pm$ 9.35        	& 58.70 $\pm$ 5.02          & 148.20 $\pm$ 14.18 			& 125.80 $\pm$ 11.00            &                    			\\\hline
\end{tabular}
\end{center}
\begin{flushleft}
$^*$Corrected for extinction using \cite{1989ApJ...345..245C} and \cite{2005ApJ...619..931I} extinction laws.
\\$^{**}$20\% uncertainties \citep{2003sws..bookQ....L, 2009A&A...500..763J}
\end{flushleft}
\end{table*}

\section{Free parameters used during SED Modeling}
\begin{table*}
\begin{center}
\caption{Free parameters used during SED Modeling.}\label{freepar}
\begin{tabular}{lll}

\hline
\hline
\textbf{Model}                          & \textbf{Free parameters} 										& \textbf{Values}                            		\\\hline
Delayed SF History 				& 1) e-folding time [{\rm Myr}] of main stellar population          	& 50, 500, 1000, 1500, 2000, 2500               	\\
                               	& 2) Age [{\rm Myr}] of the oldest stars in the galaxy          		& 500, 1000, 1500, 2000, 2500,  					\\
                                                              &                                 		& 3000, 3500, 4000, 4500, 5000 						\\\hline
Stellar Emission 		        & 1) Metallicity                                                		& 0.02, 0.05                              			\\
                               	& 2) Age [{\rm Myr}] of the separation between the  					& 5, 10, 15                         				\\
                               	& young and the old star populations              						&                         							\\\hline
Dust Attenuation               	& 1) E(B-V), the colour excess of the stellar       					& 0.10, 0.15, 0.20, 0.25, 0.30, 0.35, 0.40, 0.45,   \\
				               	& continuum light for the young population	       						&  0.50, 0.55, 0.60, 0.70              					\\
                               	& 2) Reduction factor for the E(B-V)* of the      						& 0.10, 0.15, 0.20, 0.25, 0.30, 0.35, 0.40, 0.45  	\\
                              	& old population compared to the young one       						& 0.50,	0.60, 0.70	                					\\
                               	& 3) Slope delta of the power law  										& 0.00, 0.15, 0.25, 0.40, 0.50                      \\
                                & modifying the attenuation curve                           			&                     								\\\hline
Dust Emission                  	& 1) Mass fraction of PAH [\%]                                      	& 2.50, 3.19, 3.90, 4.58, 5.26, 5.95, 6.63, 7.32    \\
                               	& 2) Minimum radiation field, $U_\mathrm{min}$ [Habing]             	& 0.1, 0.3, 0.5, 0.7, 0.8, 1.0, 1.5, 1.7, 2.0, 2.5, \\
                                & 																		&  3.0, 3.5, 4.0, 5.0, 8.0, 10.0 					\\                             
                               	& 3) power-law slope $dU/dM$\, $\propto$\,$U^\alpha$, $\alpha$      	& 1.5, 1.8, 2.0, 2.3, 2.5, 2.7, 3.0      			\\
                               	& 4) Fraction illuminated from $U_\mathrm{min}$ to $U_{max}$, $\gamma$  & 0.1, 0.2, 0.3, 0.4, 0.5, 0.6, 0.7, 0.8            \\\hline
Non-Thermal Radio Emission     	& 1) FIR/radio correlation coefficient                    				& 2.50 to 2.75                     					\\\hline                                 
AGN Torus Emission             	& 1) Ratio of the maximum to minimum                                 	& 30.0, 60.0, 100.0                 				\\
            					& radii of the dust torus                               				&                  									\\
                               	& 2) Optical depth at 9.7 microns                                       & 0.1, 1.0, 3.0, 10.0               				\\
                               	& 3) Fraction of AGN contribution to IR                                 & 0.0, 0.1, 0.2, 0.3, 0.4, 0.5, 0.6, 0.7 			\\
                               	& 4) Full opening angle of the dust torus [degree]                      & 60, 100, 140                      				\\
                               	& 5) $\beta$, from power-law density distribution for 					& -0.5, -1.0, 0.0                   				\\
                               	& the radial component of the dust torus 								&  													\\
                               	& 6) $\gamma$, from power-law density distribution for  				& 0.0, 2.0  										\\
                               	& the polar component of the dust torus 								&                         							\\\hline
\end{tabular}
\end{center}
\end{table*}


\bsp	
\label{lastpage}

\end{document}